\documentclass{aa}  

\usepackage{graphicx}
\usepackage{color}
\usepackage{natbib}
\usepackage{savefnmark}
\bibpunct{(}{)}{;}{a}{}{,}
\usepackage{txfonts}
\usepackage{arydshln}
\usepackage{dutchcal}
\usepackage{titlesec}

\titleclass{\subsubsubsection}{straight}[\subsection]

\newcounter{subsubsubsection}[subsubsection]
\renewcommand\thesubsubsubsection{\thesubsubsection.\arabic{subsubsubsection}}

\titleformat{\subsubsubsection}
  {\normalfont\normalsize}{\thesubsubsubsection}{1em}{}
\titlespacing*{\subsubsubsection}
{0pt}{3.25ex plus 1ex minus .2ex}{1.5ex plus .2ex}

\makeatletter
\def\toclevel@subsubsubsection{4}
\def\l@subsubsubsection{\@dottedtocline{4}{7em}{4em}}
\makeatother

\begin{document} 

\def\sun{\hbox{$\odot$}}
\def\degr{\hbox{$^\circ$}}
\def\arcmin{\hbox{$^\prime$}}
\def\arcsec{\hbox{$^{\prime\prime}$}}

   \title{Deep ALMA imaging of the merger NGC~1614}

   \subtitle{Is CO tracing a massive inflow of non-starforming gas?}

   \author{S. K\"onig
          \inst{1}
          \and
          S. Aalto
	   \inst{1}
          \and
          S. Muller
	   \inst{1}
	  \and
          J.~S. Gallagher III
	   \inst{2,1}
	  \and
          R.~J. Beswick
	   \inst{3}
          \and
          C.~K. Xu
           \inst{4}
          \and
          A. Evans
	   \inst{5,6}
          }

    \offprints{S. K\"onig}

   \institute{Chalmers University of Technology, Department of Earth and Space Sciences, Onsala Space Observatory, 43992 
	     Onsala, Sweden\\
              \email{sabine.koenig@chalmers.se}
         \and
             Department of Astronomy, University of Wisconsin, 475 N. Charter Street, Madison, WI, 53706, USA
         \and
             University of Manchester, Jodrell Bank Centre for Astrophysics, Oxford Road, Manchester, M13 9PL, UK
         \and
             Infrared Processing and Analysis Center, MS 100-22, California Institute of Technology, Pasadena, CA, 91125, USA
         \and
             NRAO, 520 Edgemont Road, Charlottesville, VA, 22903, USA
         \and
             University of Virginia, Charlottesville, VA, 22904, USA
             }

   \date{Received ; accepted }

 
  \abstract
   {}
   {Observations of the molecular gas over scales of $\sim$0.5 to several kpc provide crucial information on how molecular gas moves 
    through galaxies, especially in mergers and interacting systems, where it ultimately reaches the galaxy center, accumulates, and feeds nuclear 
    activity. Studying the processes involved in the gas transport is one of the important steps forward to understand galaxy evolution.}
   {$^{\rm 12}$CO, $^{\rm 13}$CO\,1$-$0, and C$^{\rm 18}$O high-sensitivity ALMA observations ($\sim$4\arcsec\,$\times$\,2\arcsec) were used to assess the 
    properties of the large-scale molecular gas reservoir and its connection to the circumnuclear molecular ring in the merger NGC~1614. Specifically, the 
    role of excitation and abundances were studied in this context. We also observed the molecular gas high-density tracers CN and CS.}
   {The spatial distributions of the detected $^{\rm 12}$CO\,1$-$0 and $^{\rm 13}$CO\,1$-$0 emission show significant differences. $^{\rm 12}$CO traces 
    the large-scale molecular gas reservoir, which is associated with a dust lane that harbors infalling gas, and extends into the southern tidal tails. 
    $^{\rm 13}$CO emission is -- for the first time -- detected in the large-scale dust lane. In contrast to $^{\rm 12}$CO, its line emission peaks 
    between the dust lane and the circumnuclear molecular ring. A $^{\rm 12}$CO-to-$^{\rm 13}$CO\,1$-$0 intensity ratio map shows high values in the ring 
    region ($\sim$30) typical for the centers of luminous galaxy mergers and even more extreme values in the dust lane ($>$45). Surprisingly, we do 
    not detect C$^{\rm 18}$O emission in NGC~1614 -- but we do observe gas emitting the high-density tracers CN and CS.}
   {We find that the $^{\rm 12}$CO-to-$^{\rm 13}$CO\,1$-$0 line ratio in NGC~1614 changes from $>$45 in the 2~kpc dust lane to $\sim$30 in the 
    starburst nucleus. This drop in ratio with decreasing radius is consistent with the molecular gas in the dust lane being kept in a diffuse, unbound 
    state while it is being funneled towards the nucleus. This also explains why there are no (or very faint) signs of star formation in the dust lane, 
    despite its high $^{\rm 12}$CO-luminosity. In the inner 1.5~kpc, the gas is compressed into denser and likely self-gravitating clouds (traced by CN and 
    CS emission), allowing it to power the intense central starburst. We find a high $^{\rm 16}$O-to-$^{\rm 18}$O abundance ratio in the 
    starburst region ($\geq$900), typical of quiescent disk gas. This is surprising since, by now, the starburst is expected to have enriched the nuclear 
    interstellar medium in $^{\rm 18}$O relative to $^{\rm 16}$O. We suggest that the massive inflow of gas may be partially responsible for the low 
    $^{\rm 18}$O/$^{\rm 16}$O abundance since it will dilute the starburst enrichment with unprocessed gas from greater radial distances. The 
    $^{\rm 12}$CO-to-$^{\rm 13}$CO abundance of $>$90 we infer from the line ratio is consistent with this scenario. It suggests that the nucleus of 
    NGC~1614 is in a transient phase of its evolution where the starburst and the nuclear growth is still being fuelled by returning gas from the minor 
    merger event.}


\titlerunning{Deep ALMA imaging of NGC~1614}

   \maketitle
%

\section{Introduction} \label{sec:intro}

\begin{table*}[ht]
\begin{minipage}[!h]{\textwidth}
\centering
\renewcommand{\footnoterule}{}
\caption{\small
 Properties of the observed lines and resulting images.}
\label{tab:obs}
\tabcolsep0.1cm
\begin{tabular}{lcccccll}
\noalign{\smallskip}
\hline
\noalign{\smallskip}
\hline
\noalign{\smallskip}
observable &  & observing frequency & beam & channel width      & sensitivity & integrated intensity & peak flux\\
\noalign{\smallskip}
           &  & [GHz]               &      & [km\,s$^{\rm -1}$] & [mJy\,beam$^{\rm -1}$] &  & [Jy]\\
\noalign{\smallskip}
\hline
\noalign{\smallskip}
continuum                   & & 105.0 & 4.\!\!\arcsec46\,$\times$\,1.\!\!\arcsec68 & -- & 0.06 & 12.9$\pm$0.8~mJy & -- \\
$^{\rm 12}$CO\,1$-$0        & & 113.5 & 4.\!\!\arcsec38\,$\times$\,1.\!\!\arcsec84 & 10   & 1.6 & 241$\pm$1.0~Jy\,km\,s$^{\rm -1}$ & 1.1$\pm$0.02 \\
CN\,1$-$0 $J$\,=\,3/2$-$1/2 & & 111.4 & 4.\!\!\arcsec06\,$\times$\,1.\!\!\arcsec61 & 10   & 1.3 & 3.8$\pm$1.0~Jy\,km\,s$^{\rm -1}$ & 0.017$\pm$0.003 \\
CN\,1$-$0 $J$\,=\,1/2$-$1/2 & & 111.7 & 4.\!\!\arcsec06\,$\times$\,1.\!\!\arcsec61 & 10   & 1.3 & 7.2$\pm$1.0~Jy\,km\,s$^{\rm -1}$ & 0.027$\pm$0.003\\
\noalign{\smallskip}
\hdashline
\noalign{\smallskip}
$^{\rm 13}$CO\,1$-$0        & & 108.5 & 4.\!\!\arcsec18\,$\times$\,2.\!\!\arcsec24 & 10   & 1.4 & 6.6$\pm$1.0~Jy\,km\,s$^{\rm -1}$ & 0.04$\pm$0.004 \\
C$^{\rm 18}$O\,1$-$0        & & 108.1 & 4.\!\!\arcsec20\,$\times$\,2.\!\!\arcsec25 & 60   & 0.8 & $<$0.1~Jy\,km\,s$^{\rm -1}$ & -- \\
CS\,2$-$1                   & &  96.4 & 4.\!\!\arcsec65\,$\times$\,2.\!\!\arcsec55 & 30   & 0.8 & 1.3$\pm$1.0~Jy\,km\,s$^{\rm -1}$ & 0.008$\pm$0.001\\
\noalign{\smallskip}
\hline
\end{tabular}
\end{minipage}
\end{table*}

Minor mergers, which are close galaxy interactions where the partners have an unequal mass ratio, constitute the majority of interacting events in the 
Universe. The gas accretion and nuclear feeding mechanisms of these mergers are different from major mergers \citep[equal-mass interactions, e.g.,][]{naa09}. 
Minor mergers are important for our understanding of galaxy evolution. In particular, they provide insights into how high-redshift spheroidal galaxies evolve 
into systems at low-redshift, such as local elliptical galaxies, and the formation of galaxy bulges \citep[e.g.,][]{naa09,wei09}. In minor mergers, gas 
introduced into the interacting system by the disturbing companion can generally be found at large radii from the center of the merger remnant 
\citep{bou05}. Stellar bars, causing gravitational torques to affect the gas, and/or tidal torques invoked by the disturbing companion, may lead to the 
transport of the molecular gas along the large-scale dust lanes to the mergers center, with this material acting to fuel both star formation and nuclear 
accretion \citep[e.g.,][]{sim80,sco85,jog06,wei09}. There the gas may form polar rings that can appear as dust lanes when seen edge-on 
\citep[e.g.,][]{com88,shl89}. We have found evidence that gas is funneled along polar rings \citep[][]{aalto00,aalto10,koenig13,koenig14}. Tracing this 
gas is vital to understand the underlying galaxy-evolution and star-formation mechanisms in this class of mergers.\\
\indent
Spectral lines of molecular isotopologs (isotopic variants) can be used to study the effect of infall, enrichment and gas physical conditions. 
Elevated line ratios between the $^{\rm 12}$C and $^{\rm 13}$C variants of CO have, for example, been proposed to indicate a replenishment of relatively 
unprocessed disk gas \citep[e.g.,][]{cas92,hen93,tani99}. However, the ratio cannot be used to infer values on $^{\rm 12}$C/$^{\rm 13}$C without taking 
effects of optical depth and line excitation (temperature, line width) into account. For example, $^{\rm 12}$CO/$^{\rm 13}$CO line ratios may be low in 
quiescent disk gas and higher in hot nuclear gas without a change in $^{\rm 12}$C/$^{\rm 13}$C abundances, or even with the opposite abundance
gradient \citep[e.g.,][]{aalto95,aalto97,dav15}. Dynamical effects are also found to have a strong influence on the observed $^{\rm 12}$CO/$^{\rm 13}$CO line 
ratios \citep[e.g.,][]{tos02,mei04,aalto10}. In addition, issues of selective photo destruction also need to be addressed \citep[e.g.,][]{aalto95}.\\
\indent
Isotopic variants of oxygen, $^{\rm 16}$O and $^{\rm 18}$O, are used to trace enrichment by massive stars \citep[e.g.,][]{hen93,gon14,fal15}. 
\citeauthor{gon14}, for example, found very high $^{\rm 18}$O abundances in the nearby ULIRG quasar \object{Mrk~231} which they attributed to an evolved 
nuclear starburst.\\
\indent
NGC~1614 \citep[D\,=\,64~Mpc, 1\arcsec\,=\,310~pc, L$_{\rm 8-1000\mu m}$ $\sim$4\,$\times$\,10$^{\rm 11}$~L$_{\sun}$,][]{san03} is a minor merger 
that shows intense nuclear activity and a complex morphology, with a minor axis dust lane crossing the optical body close to the nucleus. The 
presence of a circumnuclear ring ($r$\,$\sim$300~pc) has been reported from different tracers, for example Pa$\alpha$ \citep{alo01}, PAH 
\citep{vai12}, multiple transitions of CO at mm wavelengths \citep{koenig13,sli14,xu15}, and in the radio continuum \citep[e.g.,][]{ols10,koenig13}. Most 
of the nuclear activity originates from a very young starburst residing in the ring itself \citep[6--7~Myr,][]{pux99,koti01,schw04} and an older 
starburst at its center \citep[$>$\,10~Myr,][]{alo01}. Two competing formation scenarios for the ring have been proposed: a ``wildfire''scenario 
\citep{alo01} where the ring is the result of an outward propagating starburst, or that the ring is formed at the location of a Lindblad resonance 
and is fueled by gas moving in along the dust lanes onto the ring \citep{koenig13}.\\
\indent
A number of emission lines tracing different physical conditions originate from the molecular gas in NGC~1614. A considerable fraction of the 
total molecular gas resides in the central dust lanes but is not involved in the nuclear activity \citep[$^{\rm 12}$CO\,1$-$0,][]{ols10,sli14,gar15}.  
Further studies of the $^{\rm 12}$CO\,1$-$0 have revealed a molecular outflow in NGC~1614 \citep{gar15}.\\ 
\indent
In our previous high-resolution CO studies of NGC~1614 we not only discovered molecular gas in the starburst ring traced by CO\,2$-$1, but also 
that the majority of the CO emission of NGC~1614 is located in the dust lane and is apparently not star forming. We suggested that the gas here may be 
infalling and in a diffuse, unbound state.\\
\indent
To investigate the nature of the large-scale molecular gas of the nearby merger NGC~1614 we obtained high-sensitivity ALMA $^{\rm 12}$CO, $^{\rm 13}$CO, 
and C$^{\rm 18}$O\,1$-$0 data. We also detected CN\,1$-$0 and CS\,2$-$1.\\
\indent
In this paper we present the results of a study of the properties of the molecular gas within a radius of $\sim$3~kpc in NGC~1614. An analysis of our 
higher angular resolution ALMA data will be presented in a later paper.Throughout the paper, we are concerned with pure rotational transitions of the 
observed molecules between upper state $J'$\,=\,$j$ and lower state $J$\,=\,$i$ that are labeled $j$\,$-$\,$i$. Section\,\ref{sec:obs} describes the 
observations and the data analysis, in Sect.\,\ref{sec:results} the results of the observations are presented, and we discuss their implications 
in Sect.\,\ref{sec:discussion}.


\section{Observations, data reduction and analysis} \label{sec:obs}

\begin{figure}[t]
  \centering
    \includegraphics[width=0.4\textwidth]{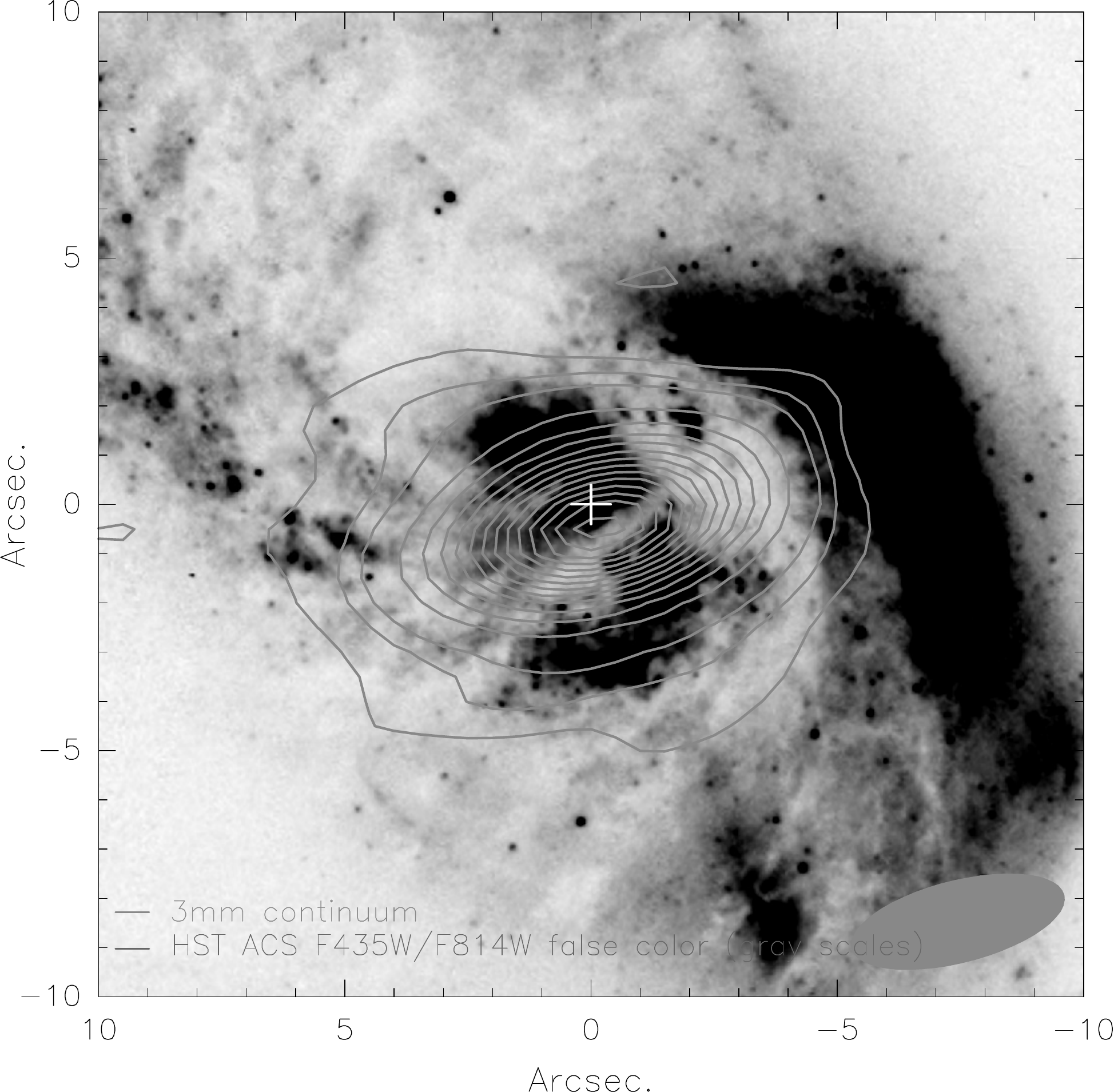}
  \caption{\footnotesize 3~mm continuum emission (in gray contours) on top of an \textit{HST} ACS F435W filter gray-scale image to facilitate a 
   comparison of the location and distribution of the emission with respect to the dust lanes at the center of NGC~1614. This image of the 3~mm 
   continuum emission was created using uniform weighting, which resulted in a sensitivity of 1$\sigma$\,=59.8~$\mu$Jy\,beam$^{\rm -1}$. The contours 
   start at 3$\sigma$ and are spaced in steps of 5$\sigma$. The cross marks the phase center of the observations, placed at the center of the molecular 
   gas ring. The beam (see also Table\,\ref{tab:obs}) is shown in the lower right corner. North is up, east to the left.}
  \label{fig:overlay_cont_false_color}
\end{figure}
\begin{figure*}[ht]
  \begin{minipage}[hbt]{0.4925\textwidth}
  \centering
    \includegraphics[width=0.75\textwidth]{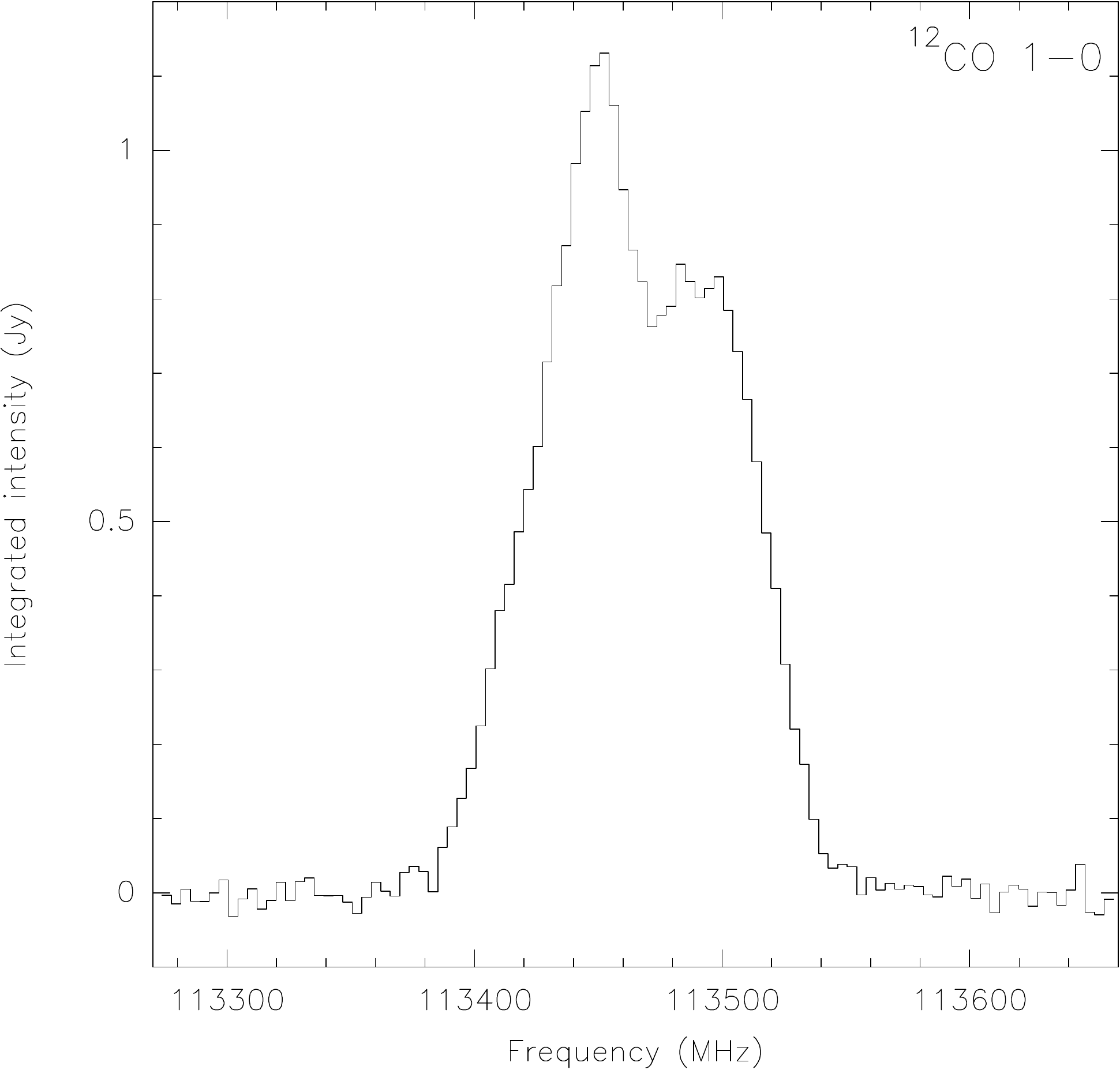}
  \end{minipage}
  \begin{minipage}[hbt]{0.4925\textwidth}
  \centering
    \includegraphics[width=0.75\textwidth]{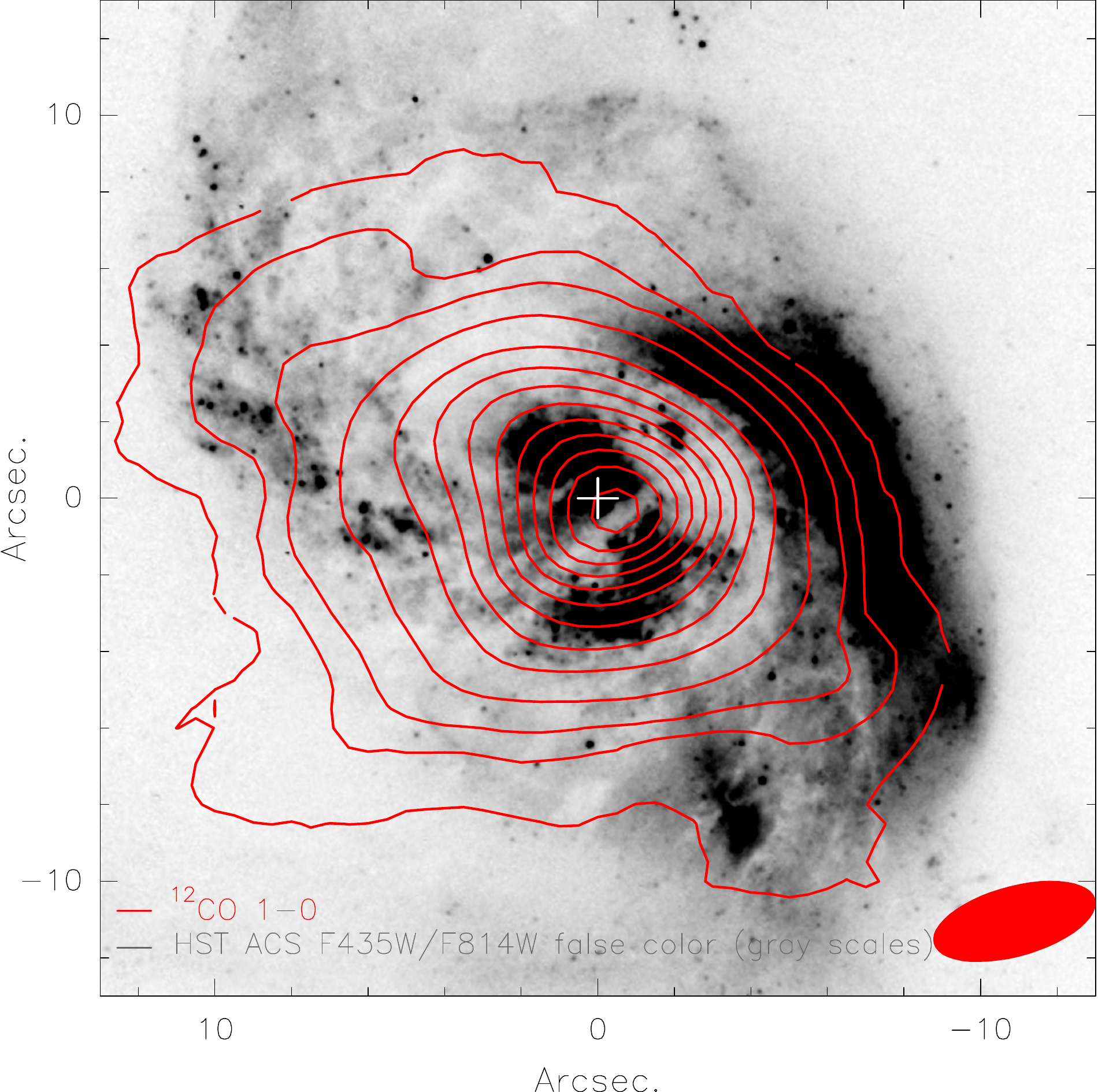}
  \end{minipage}
  \caption{\footnotesize $^{\rm 12}$CO\,1$-$0 emission in NGC~1614. The spectrum of the $^{\rm 12}$CO\,1$-$0 emission with a channel width of 
   10~km\,s$^{\rm -1}$ is shown on the \textit{left}, an overlay of the integrated emission contours on top of an \textit{HST} ACS F435W filter 
   gray-scale image is shown on the \textit{right}. The integrated intensity map was obtained using uniform weighting. The resulting 1$\sigma$ 
   sensitivity is $\sim$1.6~mJy\,beam$^{\rm -1}$, the synthesized beam (see Table\,\ref{tab:obs}) is depicted in the lower right corner. The first 
   three contour levels are at 3$\sigma$, 13$\sigma$, and 23$\sigma$. For legibility purposes, the next contour is at 50$\sigma$; from there on the 
   subsequent contours are spaced by 50$\sigma$ until the peak flux is reached. The cross marks the phase center. North is up, east to the left.}
  \label{fig:overlay_12co_false_color}
\end{figure*}
\begin{figure*}[h]
  \begin{minipage}[hbt]{0.4925\textwidth}
  \centering
    \includegraphics[width=0.75\textwidth]{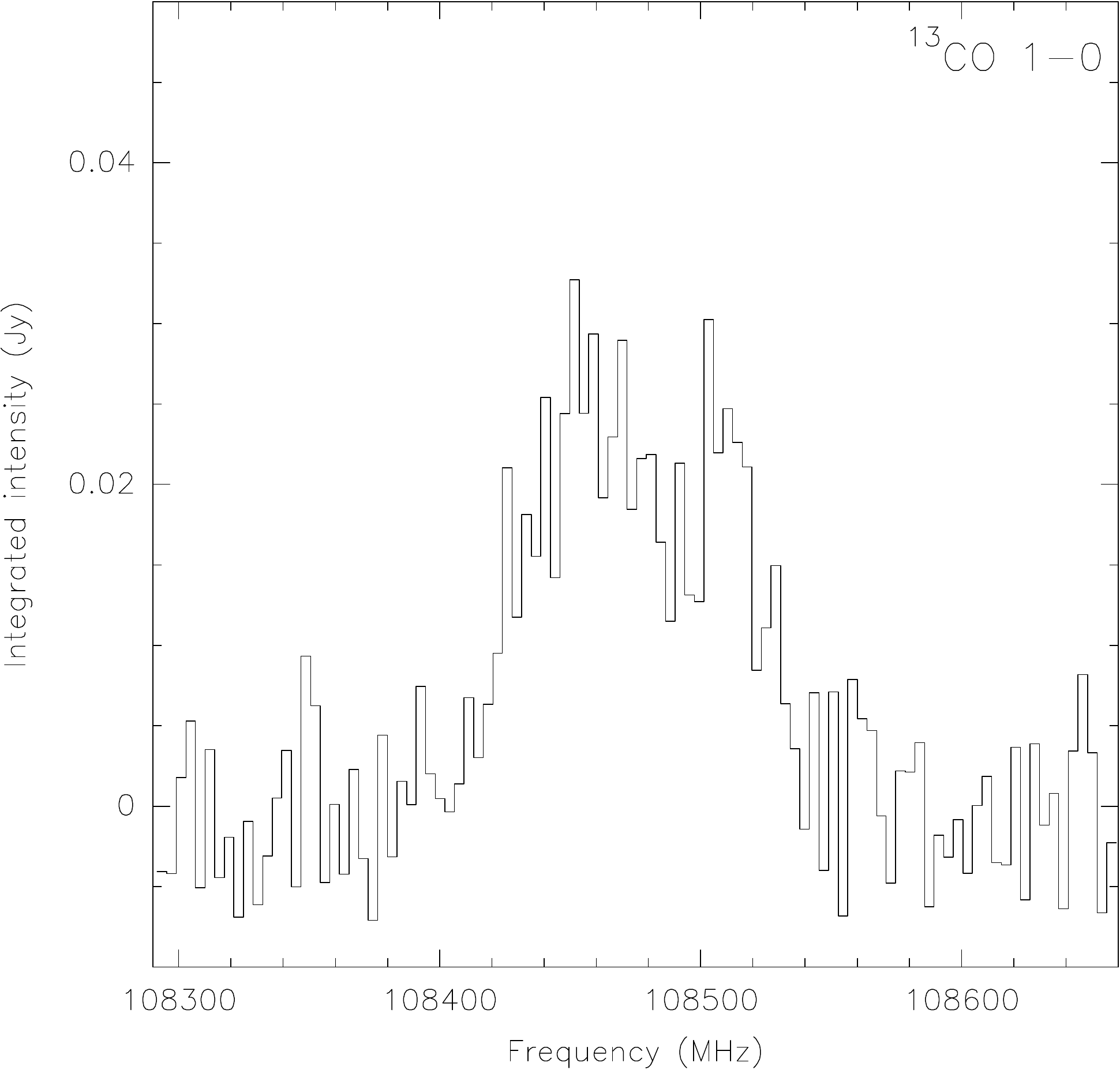}
  \end{minipage}
  \begin{minipage}[hbt]{0.4925\textwidth}
   \begin{center}
    \includegraphics[width=0.75\textwidth]{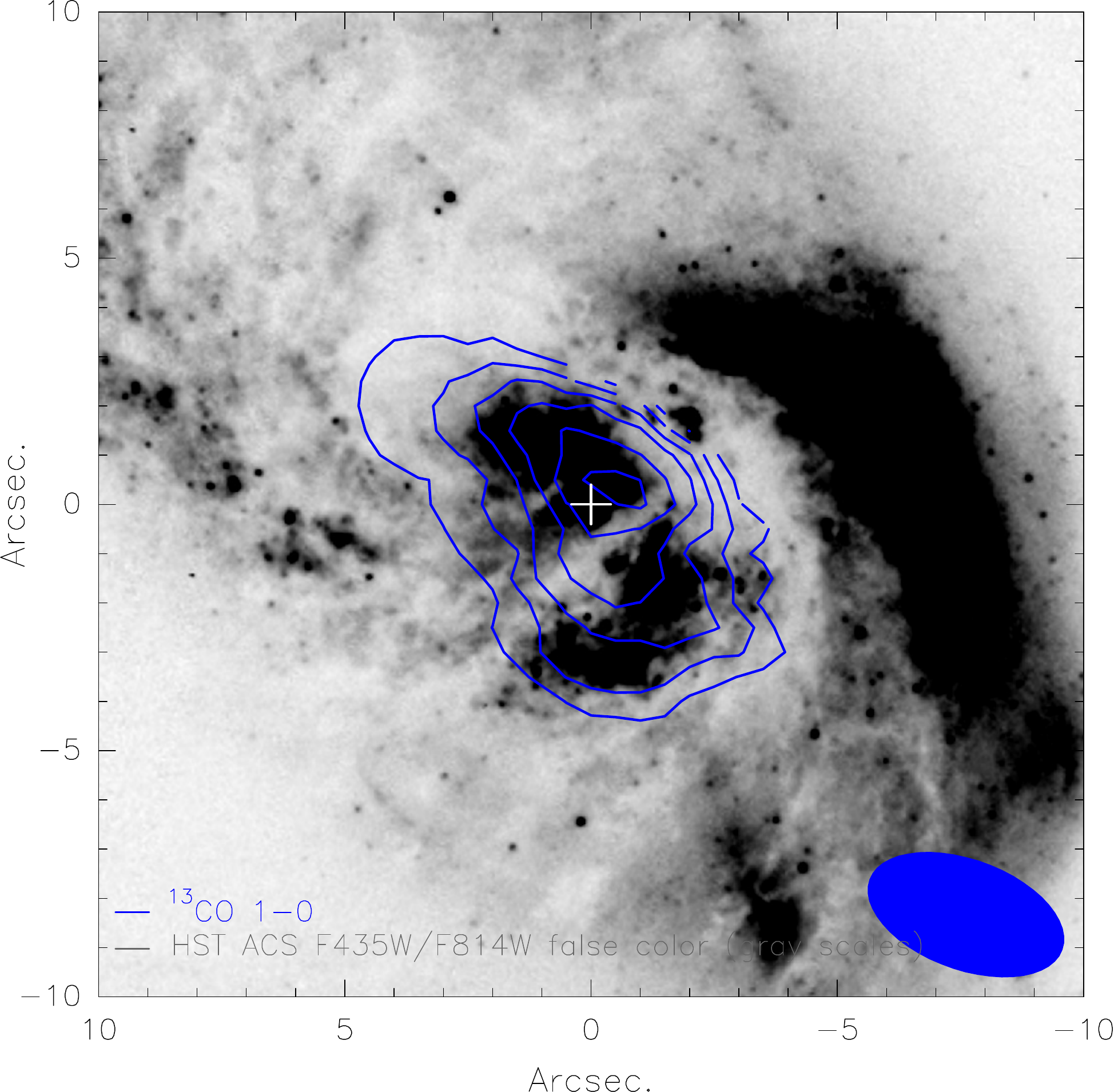}
   \end{center}
  \end{minipage}
  \caption{\footnotesize Spectrum of the $^{\rm 13}$CO\,1$-$0 emission in NGC~1614 (channel width: 10~km\,s$^{\rm -1}$, \textit{left}), and overlay of 
   the integrated emission contours on top of an \textit{HST} ACS F435W filter gray-scale image (\textit{right}). The contours in the map 
   obtained from natural weighting start at 4$\sigma$ and are spaced in steps of 3$\sigma$. The 1$\sigma$ sensitivity here is 
   $\sim$1.4~mJy\,beam$^{\rm -1}$. The beam (see Table\,\ref{tab:obs}) is depicted in the bottom right corner. The cross marks the phase center. North 
   is up, east to the left.}
  \label{fig:overlay_13co_false_color}
\end{figure*}
\begin{figure*}[h]
  \begin{minipage}[hbt]{0.4925\textwidth}
  \centering
    \includegraphics[width=0.75\textwidth]{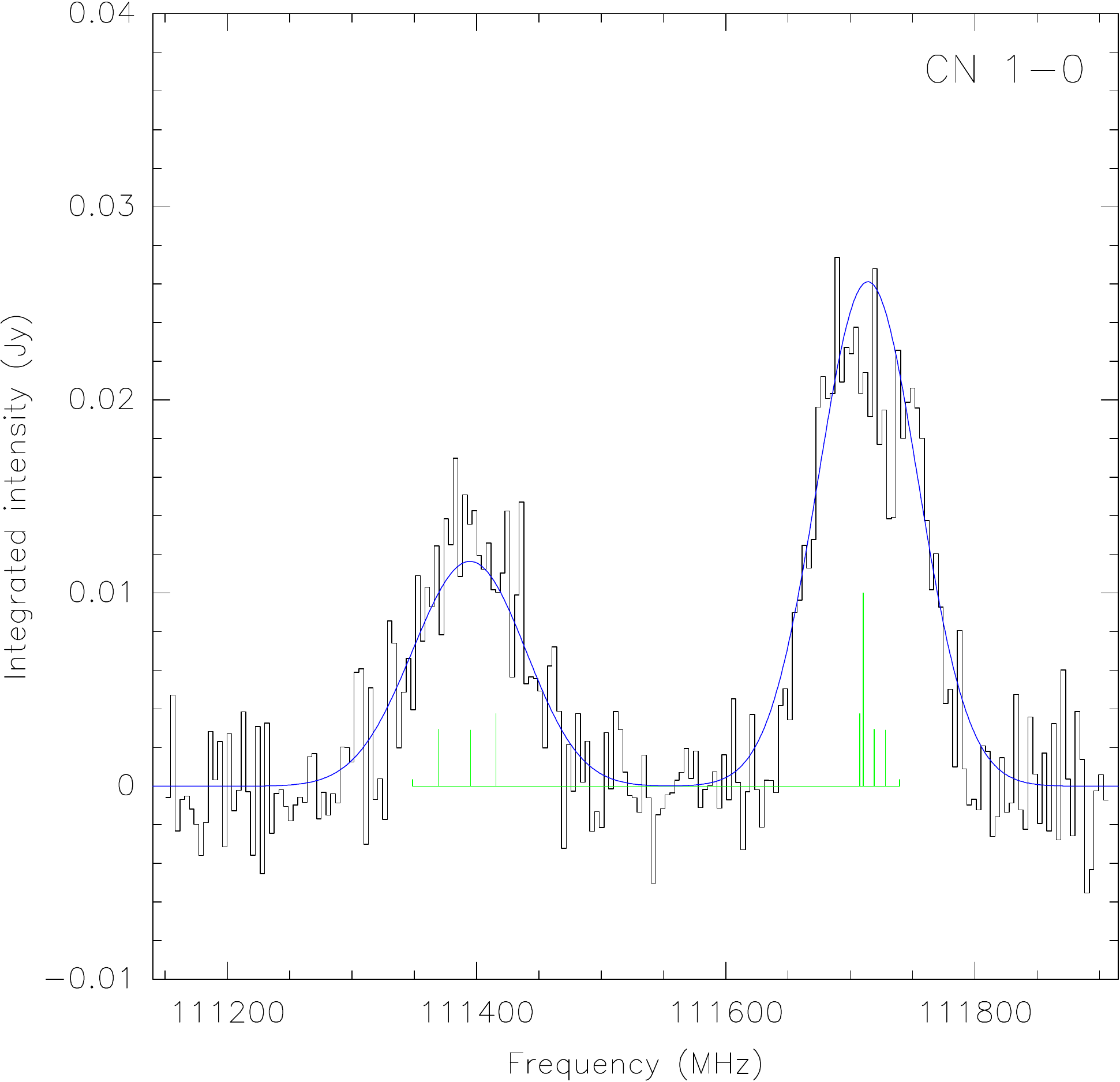}
  \end{minipage}
  \begin{minipage}[hbt]{0.4925\textwidth}
  \centering
    \includegraphics[width=0.75\textwidth]{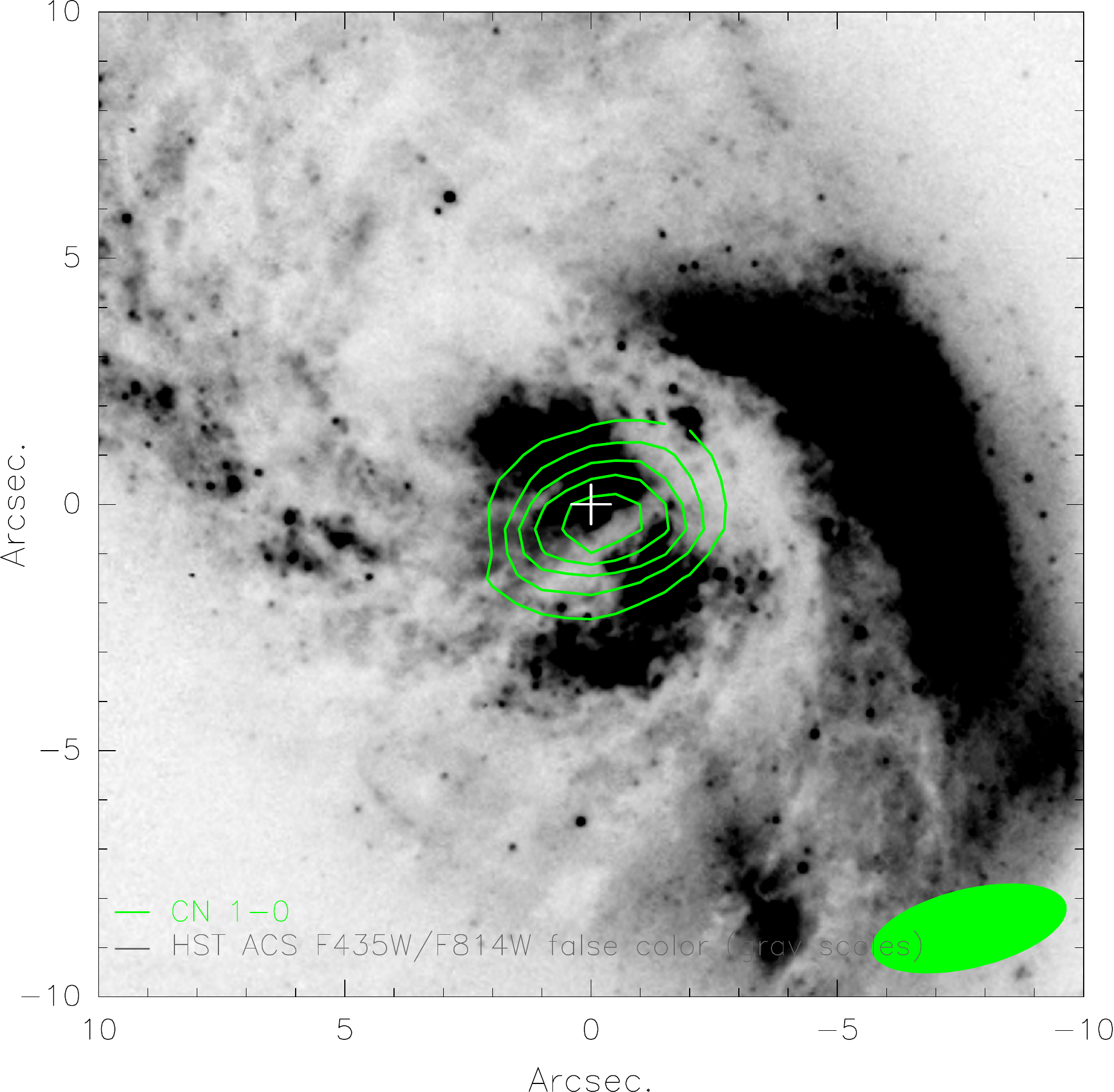}
  \end{minipage}
  \caption{\footnotesize Spectrum of the CN\,1$-$0 emission in NGC~1614 (\textit{left}), and overlay of the integrated emission on top of an 
   \textit{HST} ACS F435W filter gray-scale image (\textit{right}). In the spectrum (channel width: 10~km\,s$^{\rm -1}$) the hyperfine-structure 
   components of CN\,1$-$0 are indicated in green. The blue line represents the convolution of the CN hyperfine structure with a Gaussian profile. This 
   indicates that the gas traced by the CN\,1$-$0 emission in NGC~1614 is optically thin. The integrated intensity map was obtained using uniform 
   weighting. The resulting sensitivity is $\sim$1.3~mJy\,beam$^{\rm -1}$, the resulting beam is depicted in the bottom right corner. The contour 
   levels of the integrated emission start at 5$\sigma$ and are spaced in steps of 5$\sigma$. The beam (see also Table\,\ref{tab:obs}) is depicted in 
   the lower right corner. The cross marks the phase center. North is up, east to the left.}
  \label{fig:cn_spec+map}
\end{figure*}
\begin{figure*}[h]
  \begin{minipage}[hbt]{0.4925\textwidth}
  \centering
    \includegraphics[width=0.75\textwidth]{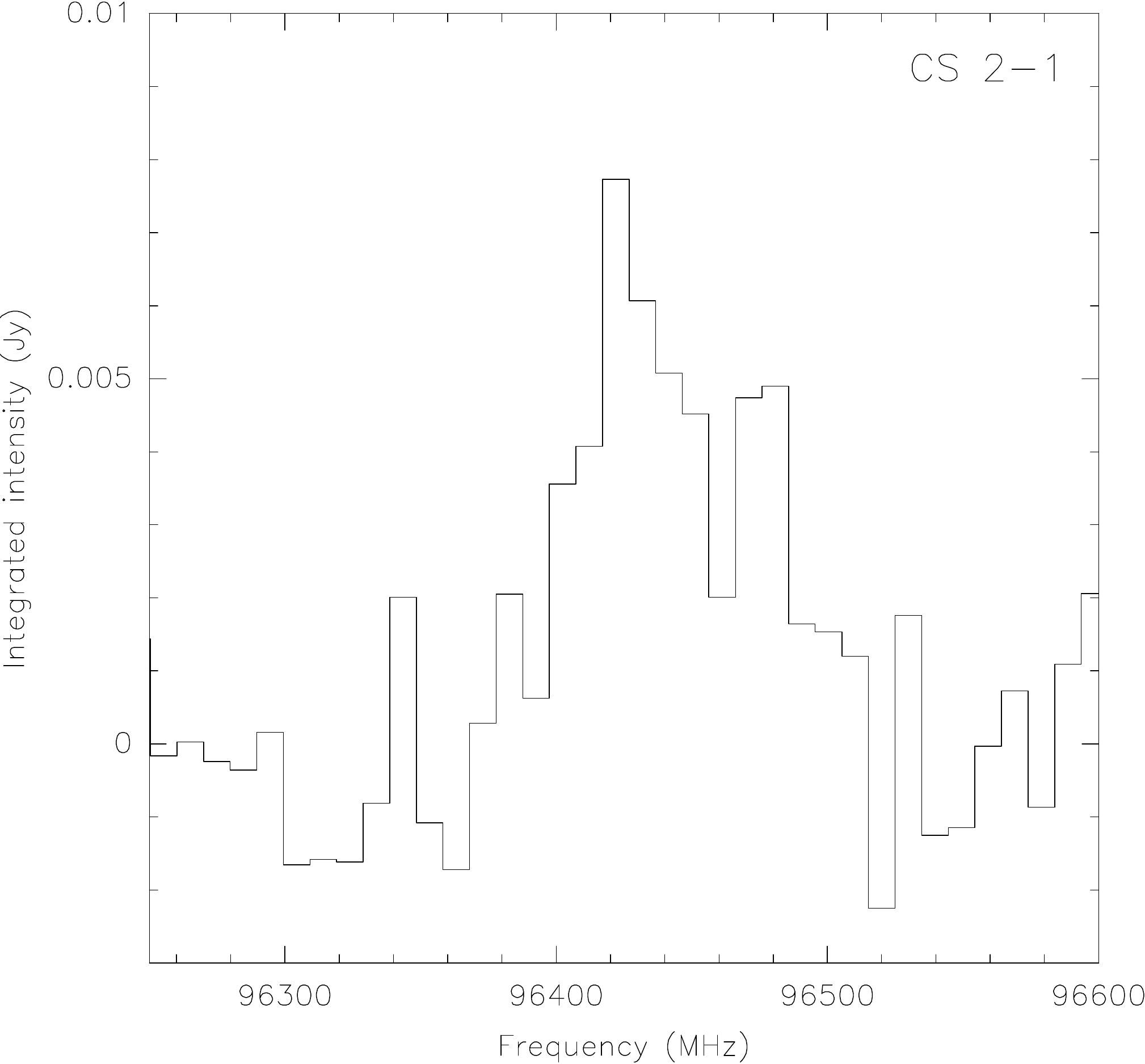}
  \end{minipage}
  \begin{minipage}[hbt]{0.4925\textwidth}
  \centering
    \includegraphics[width=0.75\textwidth]{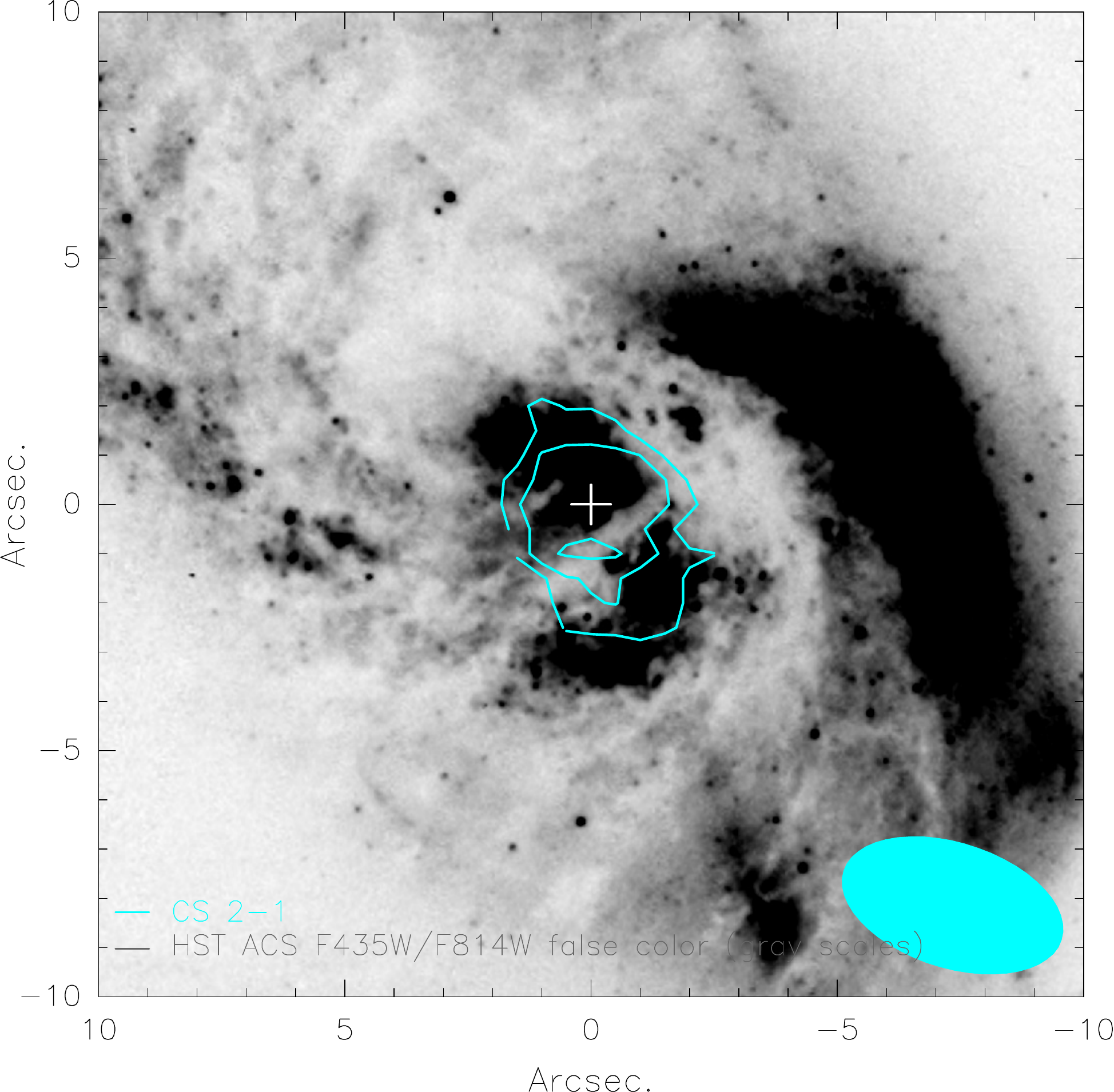}
  \end{minipage}
  \caption{\footnotesize Spectrum of the CS\,2$-$1 emission in NGC~1614 (\textit{left}), and overlay of the integrated emission on top of an 
   \textit{HST} ACS F435W filter gray-scale image (\textit{right}). The channels in the spectrum were binned to 30~km\,s$^{\rm -1}$. The contours 
   of the integrated emission in the naturally weighted map start at 3.5$\sigma$ and are spaced in steps of 3$\sigma$ 
   (1$\sigma$\,=\,0.8~mJy\,beam$^{\rm -1}$). The beam (see also Table\,\ref{tab:obs}) is depicted. The cross marks the phase center. North is up, east 
   to the left.}
  \label{fig:cs_spec+map}
\end{figure*}
\begin{figure*}[h]
  \begin{minipage}[hbt]{0.4925\textwidth}
  \centering
    \includegraphics[width=0.808\textwidth]{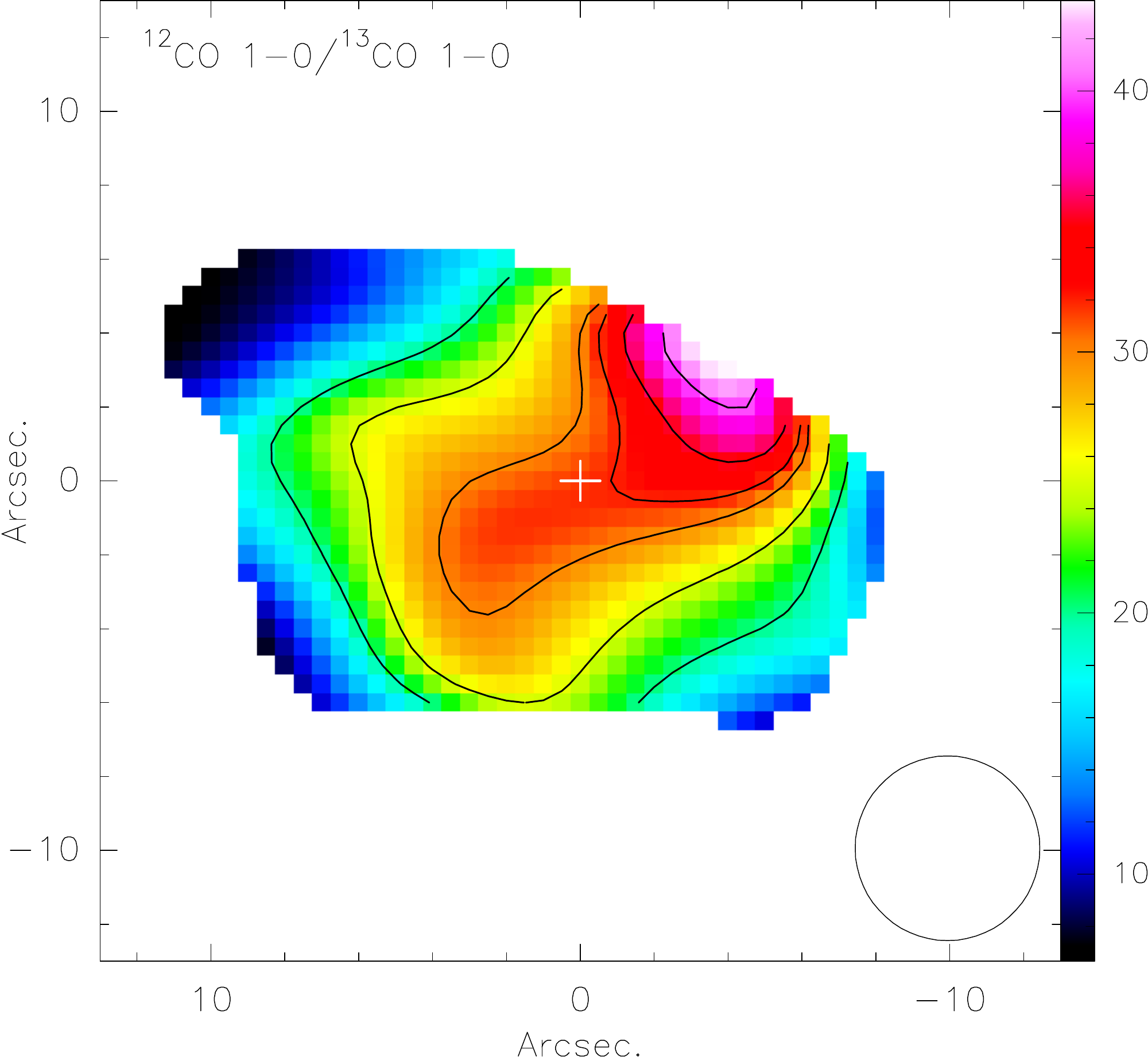}
  \end{minipage}
  \begin{minipage}[hbt]{0.4925\textwidth}
  \centering
    \includegraphics[width=0.75\textwidth]{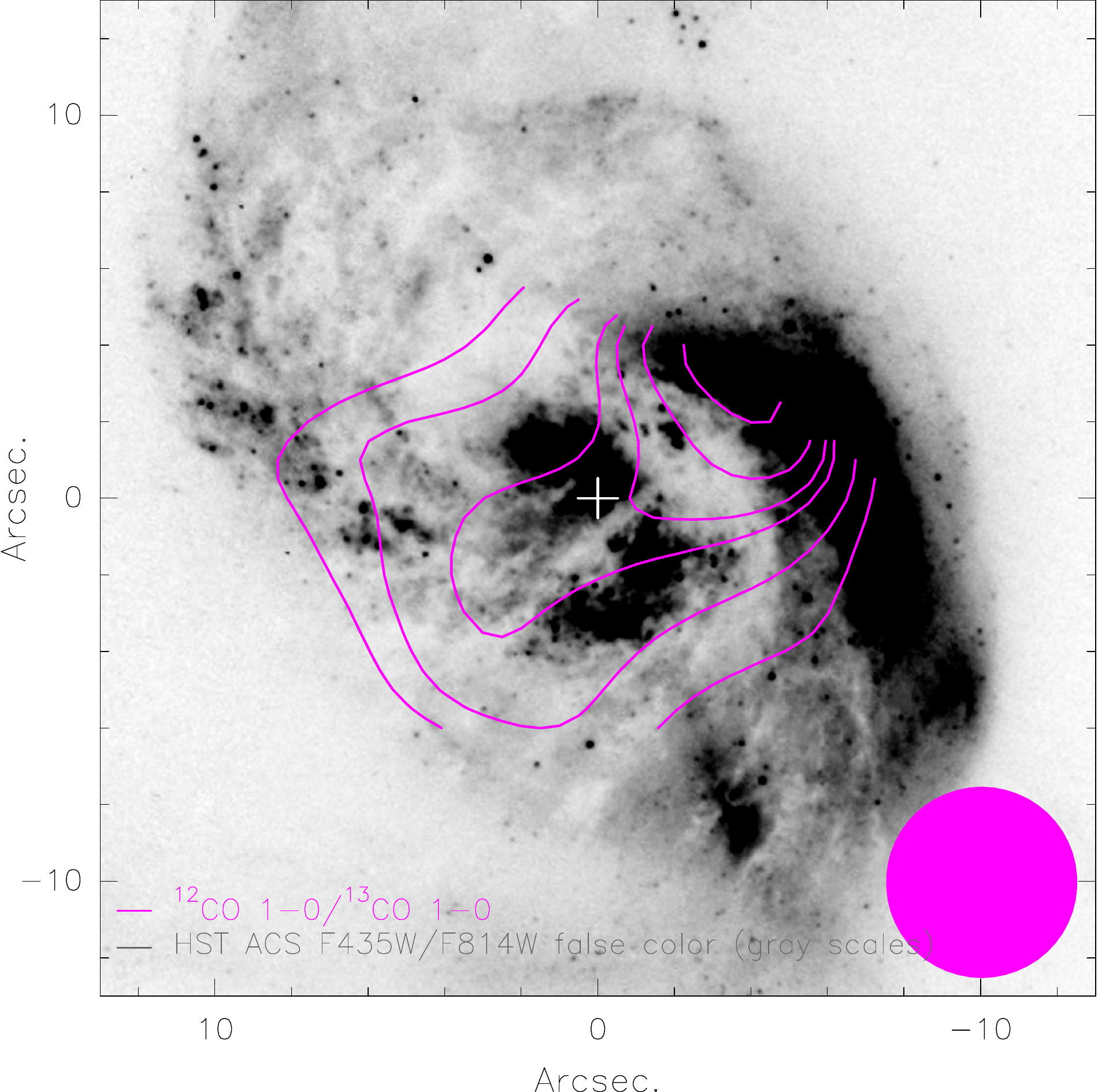}
  \end{minipage}
  \caption{\footnotesize $^{\rm 12}$CO-to-$^{\rm 13}$CO integrated intensity line ratio ($\mathcal{R_{\rm 10}}$) map (\textit{left}) and 
   overlay of the same in contours on top of an \textit{HST} ACS F435W filter gray-scale image (\textit{right}). The ratio decreases from about 
   40$-$45 at the northwestern edge down to $\sim$30 along the nuclear dust lane, and $\sim$10-20 at the northeastern, eastern, and western edges. 
   $\mathcal{R_{\rm 10}}$ seems to be quite constant along the nuclear dust lane until the ratio steeply increases toward the dust lane in the west. North 
   is up, east to the left, and the beam is shown in the lower right corner of each map. The cross marks the position of the phase center.}
  \label{fig:overlay_12co-13co-ratio_false_color}
\end{figure*}

\object{NGC~1614} was observed with the the Atacama Large Millimeter/submillimeter Array (ALMA) in band 3 (3~mm) as part of Cycle 2 observations of 
project 2013.1.00991.S (PI: S. K\"onig). Here we focus on the data obtained with the ALMA Compact Array (ACA) and the ALMA main array in compact 
configuration. The ACA observations took place on 2014 June 16 and 30, the ALMA compact configuration observations on 2014 December 4 and 5. The phase 
center of the observations was located at $\alpha$=04:34:00.02688 and $\delta$=$-$08.34.44.5692 (J2000). Two tunings were obtained centered on 113.5~GHz 
and 108.5~GHz, where the redshifted frequencies of the $^{\rm 12}$CO\,1$-$0 and $^{\rm 13}$CO\,1$-$0 lines are located. Other lines observed within the 
ALMA band are C$^{\rm 18}$O\,1$-$0, CN\,1$-$0 and CS\,2$-$1. The 3~mm continuum was reconstructed using line-free channels. We used four spectral windows 
of 1.875~GHz bandwidth ($\sim$5000~km\,s$^{\rm -1}$) each, with a velocity resolution of $\sim$5~km\,s$^{\rm -1}$, after Hanning smoothing. 
During the observations, different sources were observed for calibration purposes: \object{Uranus} as flux calibrator, \object{J0339-0146} as bandpass 
calibrator, and \object{J0423-0120} as bandpass and phase calibrator. With baseline lengths of between 8.8~m and 49.0~m, the ACA data are sensitive to 
scales smaller than $\sim$38\arcsec. The compact configuration main array ALMA data are sensitive to scales smaller than $\sim$22\arcsec\ (baselines: 
15.0~m\,$-$\,390~m).\\
\indent
For all emission lines, sets of visibilities from ACA and the compact configuration observations were combined by weighting the individual data sets in 
the uv plane to guarantee a continuous amplitude vs. uv-distance distribution. The combined data set was then deconvolved using the 
``clean'' task in CASA, and individual data cubes were created for each observable (continuum and emission lines). The resulting beam sizes and 
sensitivities are shown in Table\,\ref{tab:obs}.\\
\indent
After calibration and imaging within CASA\footnote{http://casa.nrao.edu/} \citep{mcm07}, both visibility sets were converted into FITS format and imported 
in the GILDAS/MAPPING\footnote{http://www.iram.fr/IRAMFR/GILDAS} for further analysis. All integrated intensity maps in this paper are moment-zero maps. 
To obtain the $^{\rm 12}$CO-to-$^{\rm 13}$CO\,1$-$0 map in Fig.\,\ref{fig:overlay_12co-13co-ratio_false_color} we used the integrated intensity maps of 
the two emission line distributions without a-priori clipping. The resulting maps were smoothed to a common beam at 5\arcsec\,$\times$\,5\arcsec, and then 
the ratio map was obtained on a pixel-by-pixel basis. Only pixels with signal-to-noise ratios equal to or higher than 5 in the ratio map were taken into 
account and are depicted in the resulting image.


\section{Results} \label{sec:results}

\subsection{3~mm continuum} \label{subsec:cont}

The 3~mm continuum (Fig.\,\ref{fig:overlay_cont_false_color}) is centrally peaked at the nucleus of NGC~1614. The total integrated flux recovered 
from the area enclosed by the the 3$\sigma$ contours in the uniformly weighted map is $\sim$12.9~mJy. The size and structure of the distribution, 
however, show more extended features as well. The emission seems to be elongated with a northeast-southwest direction similar to what was observed 
for $^{\rm 12}$CO\,1$-$0 previously \citep{ols10,sli14}. The continuum emission is clearly associated with the bulk of the $^{\rm 12}$CO emission, but 
also with the dust lanes in the center of NGC~1614. Previously published observations of the 1.3~mm continuum \citep{wil08,koenig13} show a more 
compact distribution ($\sim$6-7\arcsec) of the emission, even with comparable spatial resolution. This might be partially due to the excellent continuum 
sensitivity in our observations compared to previous data sets, but it could also be that the continuum at higher frequencies is more associated 
with the gas in the molecular ring \citep{koenig13} than the larger-scale molecular gas reservoir \citep{ols10,sli14}.

\subsection{$^{\rm 12}$CO\,1$-$0} \label{subsec:12co1-0}

The $^{\rm 12}$CO\,1$-$0 emission in NGC~1614 is more extended than the 3~mm continuum emission. In addition to the difference in size, the 
$^{\rm 12}$CO distribution also shows a significantly different structure than the 3~mm continuum (Fig.\,\ref{fig:overlay_12co_false_color}). 
Location and general morphology are in agreement with the results of \citet{ols10}, \citet{sli14}, and \citet{gar15}: The gas distribution appears 
centrally peaked with a significant elongation along a northeast-southwest direction, associated with the dust lanes in the center of NGC~1614. 
\citet{ols10} found an extent of about 11\arcsec, \citet{sli14} about 14\arcsec, \citet{gar15} about 12\arcsec. In our data the extension is roughly 
25\arcsec\ in the northeast-southwest direction. This difference is due to the higher sensitivity and the integration of short-spacing observations in 
our data set. Additionally, the distribution in Fig.\,\ref{fig:overlay_12co_false_color} shows features in the 
molecular gas that were previously unknown. The extension towards the south, for example, consists of low-surface brightness gas that extends into the 
tidal tails to the south-southwest of the main body of the galaxy. Hints for the feature extending toward the southeast, continuing the direction of 
the minor nuclear dust lane, can already be detected in \citet{sli14} and \citet{gar15}. Our higher-sensitivity, improved uv-coverage observations now 
clearly confirm their presence. The total interferometric integrated flux enclosed in the 3$\sigma$ contours is $\sim$241~Jy\,km\,s$^{\rm -1}$, 
which is about a factor of seven times more than in previous observations of \cite{ols10}, and a factor of two more than previously observed by 
\citet{sco89}. A comparison to single-dish observations shows that we recover approximately 115$\%$ of the flux that has recently been detected by
\citet{cos11}.\\
\indent
The spectrum of the $^{\rm 12}$CO\,1$-$0 emission (Fig.\,\ref{fig:overlay_12co_false_color}a) shows a double-peaked Gaussian with a FWHM line width 
of $\sim$250~km\,s$^{\rm -1}$. The peak flux is at about 1.1~Jy.

\subsection{$^{\rm 13}$CO\,1$-$0} \label{subsec:13co1-0}

Figure\,\ref{fig:overlay_13co_false_color} shows the spectrum and integrated intensity distribution of the $^{\rm 13}$CO\,1$-$0 emission in NGC~1614. The 
emission line appears to be double peaked, with its highest flux being $\sim$0.04~Jy, and an integrated flux of $\sim$6.6~Jy\,km\,s$^{\rm -1}$ at a line 
width (FWHM) of about 250~km\,s$^{\rm -1}$. The recovered flux corresponds to 98$\%$ of the total flux detected with single-dish observations 
\citep{cos11}. Compared to previous interferometric observations, the sensitivity in our observations is increased by a factor of seven. As a result we 
recover an integrated flux that is twice as high as in these data \citep{sli14}. The integrated intensity distribution shows the 
$^{\rm 13}$CO\,1$-$0 emission to be situated in an elongated structure of roughly 12\arcsec\ extending from the northeast of the nucleus of NGC~1614 to 
the southwest of it. The peak of the distribution, however, is located slightly northwest of the nucleus, where the $^{\rm 12}$CO\,2$-$1 emission in the 
molecular ring connects to the dust lane \citep[Fig.\,\ref{fig:overlay_12co2-1_on_13co1-0},][]{koenig13}. A comparison with an \textit{Hubble Space 
Telescope (HST)} image (Fig.\,\ref{fig:overlay_13co_false_color}b) shows that the $^{\rm 13}$CO emission appears to follow the main high brightness star 
forming structures at the center of NGC~1614. The image furthermore indicates that the extent of the $^{\rm 13}$CO emission does not cover the complete 
extent of the western dust lane. Instead, it seems that the bulk of the $^{\rm 13}$CO\,1$-$0 emission is avoiding the dust lane, in contrast to 
the $^{\rm 12}$CO\,1$-$0 emission (see also Fig.\,\ref{fig:overlay_13co1-0_on_12co1-0}).

\subsection{C$^{\rm 18}$O\,1$-$0CN\,1$-$0, and CS\,2$-$1} \label{subsec:other_lines}

Other molecular emission lines contained in the observed bandwidths are C$^{\rm 18}$O\,1$-$0, CN\,1$-$0, and CS\,2$-$1.\\
\indent
The C$^{\rm 18}$O\,1$-$0 emission line was not detected. An upper limit at 3$\sigma$ amounts to $\sim$0.1~Jy\,km\,s$^{\rm -1}$ within the central 
4\arcsec\,$\times$\,4\arcsec. This flux limit is a factor of more than 8.5 lower than what has been obtained from previous single-dish observations 
\citep{cos11}.\\
\indent
The CN\,$N$=1$-$0 emission is located in a compact, unresolved distribution at the nucleus of NGC~1614 (Fig.\,\ref{fig:cn_spec+map}bc). With a peak flux 
of $\sim$0.03~Jy and an integrated flux of 7.2~Jy\,km\,s$^{\rm -1}$, the $J$\,=\,3/2$-$1/2 component is stronger than the $J$\,=\,1/2$-$1/2 component 
(Fig.\,\ref{fig:cn_spec+map}a; peak flux: $\sim$0.02~Jy, integrated flux: $\sim$3.8~Jy\,km\,s$^{\rm -1}$). The total integrated flux is 
$\sim$10.3~Jy\,km\,s$^{\rm -1}$. The average FWHM line width amounts to $\sim$250~km\,s$^{\rm -1}$. $\mathcal{R}_{\rm CN}$, the integrated line ratio 
between the CN\,$N$\,=\,1$-$0 $J$\,=\,3/2$-$1/2 and $J$\,=\,1/2$-$1/2 components amounts to $\sim$2, which indicates that the gas traced by the CN 
emission in NGC~1614 is optically thin.\\
\indent
The CS\,2$-$1 emission is clearly detected in NGC~1614 (Fig.\,\ref{fig:cs_spec+map}). Its spatial distribution is compact and centered on the nucleus 
of the galaxy, following the main high-brightness starforming structures in the \textit{HST} image (Fig.\,\ref{fig:cs_spec+map}b). The peak flux in the 
spectrum is at 7.7~mJy. The integrated flux is found to be roughly 1.3~Jy\,km\,s$^{\rm -1}$, at a FWHM line width of $\sim$250~km\,s$^{\rm -1}$.

\subsection{$^{\rm 12}$CO-to-$^{\rm 13}$CO\,1$-$0 line ratio} \label{subsec:12co-13co_line_ratio}

From a detailed comparison of the $^{\rm 12}$CO and $^{\rm 13}$CO\,1$-$0 emission, the distributions indicate an offset in the emission peak location 
between the two tracers. The $^{\rm 12}$CO peaks exactly at the nucleus of NGC~1614, whereas $^{\rm 13}$CO peaks slightly to the northwest of it (an 
overlay image showing the central 14\arcsec\ can be found in the Appendix in Fig.\,\ref{fig:overlay_13co1-0_on_12co1-0}). Furthermore, the two emission 
distributions are distinctly different in size and structure. A $^{\rm 12}$CO-to-$^{\rm 13}$CO\,1$-$0 ($\mathcal{R_{\rm 10}}$) line ratio map 
(Fig.\,\ref{fig:overlay_12co-13co-ratio_false_color}) shows compelling evidence for this. The $\mathcal{R_{\rm 10}}$ distribution shows an elongated 
structure of constant value ($\sim$30) that coincides with the location of the minor axis dust lane close to the nucleus 
(Fig.\,\ref{fig:overlay_12co-13co-ratio_false_color}b). Moving along this direction toward the dust lane in the west of the nucleus, 
$\mathcal{R_{\rm 10}}$ increases to its highest values of 40 to 45 at the outer edge of the western dust lane, where the $^{\rm 13}$CO\,1$-$0 is largely 
absent. Toward the northeastern, eastern, and western edge of the map, the distribution shows more normal values ($\mathcal{R_{\rm 10}}$\,=\,10$-$15) 
associated with the gas in the disk of starburst galaxies.\\
\indent
\citet{sli14} reported $\mathcal{R_{\rm 10}}$ similar to our findings. Their values range from about 25 to approximately 40$-$45. Their line ratio map, 
however, shows a different distribution compared to our work (Fig.\,\ref{fig:overlay_12co-13co-ratio_false_color}) at several locations in NGC~1614: the 
highest values for $\mathcal{R_{\rm 10}}$ are found toward the southern part of the circumnuclear molecular ring. $\mathcal{R_{\rm 10}}$ is slightly lower 
than that to the west (toward the dust lane). The reason might be that \citet{sli14} only recovered $\sim$50$\%$ of the total flux in their data set,
while we essentially recovered 100\% of the single dish flux. The authors found the same lower $\mathcal{R_{\rm 10}}$ toward the northeast and east of the 
ring, however.\\
\indent
Higher resolution observations are necessary to exactly locate the peak in the $\mathcal{R_{\rm 10}}$ distribution (K\"onig et al., in prep.).


\section{Discussion} \label{sec:discussion}

\subsection{Line ratio variations} \label{subsec:line_ratios}

For the first time we were able to measure the $^{\rm 12}$CO-to-$^{\rm 13}$CO\,1$-$0 line ratio in the central dust lanes in NGC~1614 
(Sect.\,\ref{subsec:12co-13co_line_ratio}). We report an overall high $\mathcal{R_{\rm 10}}$, with increasing values toward the dust western lane. Studies 
have been published in other nearby minor merger systems, for example, the \object{Medusa merger} \citep{aalto10}, which show similarly elevated 
$\mathcal{R_{\rm 10}}$. It was suggested that the elevated line ratios are the result of two effects, either acting alone or combined: 1) changes in the 
physical conditions in the gas, or 2) abundance effects \citep[e.g.,][]{mei04,aalto10}. These effects are discussed further below.

\subsubsection{Excitation effects} \label{subsubsec:excitation}

\subsubsubsection{Temperature} \label{subsubsubsec:excitation_temperature}

In the context of the discussion in this section, we assume that the abundances of $^{\rm 12}$CO and $^{\rm 13}$CO are constant over the central 
region of interest. High average gas temperatures, low average gas densities, or exceptionally high velocity dispersions in the molecular clouds could 
be responsible for elevated line ratios \citep[e.g.,][]{aalto95,mei04}. In NGC~1614, the gas with the highest $\mathcal{R_{\rm 10}}$ ratios is located 
in the dust lane, away from the bulk of the ongoing star formation. This most likely excludes the notion of high gas temperatures as the cause of 
the elevated line ratio. An alternative mechanism to elevate $\mathcal{R_{\rm 10}}$, and the absence of star formation in the dust lane, could be 
mechanical heating and shears caused by shocks. So far, no indications for shocks in the dust lane in NGC~1614 have been found: 
$^{\rm 13}$CO\,2$-$1 observations at comparable spatial resolutions \citep{wil08} indicate an emission distribution similar to what we present for 
$^{\rm 13}$CO\,1$-$0 in this work. Assuming that the elevated $\mathcal{R_{\rm 10}}$ values are due to the presence of high-density gas, we expect 
densities of about 3\,$\times$10$^{\rm 3}$~cm$^{\rm -3}$ or more. The result would be that the $^{\rm 13}$CO\,2$-$1 flux is a factor of four or more higher 
than for $^{\rm 13}$CO\,1$-$0. This would have been picked up by the observations of \citet{wil08}. Thus, the presence of high-density gas due to shocks can 
most likely be ruled out as the determining factor for the high line ratios in NGC~1614. Furthermore, observations of dense gas tracers like HCN, 
HCO$^{\rm +}$\,4$-$3 and CO\,6$-$5 \citep{ima13,sli14} have not yielded a detection of dense gas emission in the dust lane, their emission is solely found 
in the circumnuclear ring. 

\subsubsubsection{Density} \label{subsubsubsec:excitation_density}

Change in the gas density is a valid option to cause the observed change in $\mathcal{R_{\rm 10}}$ in NGC~1614, however. A decrease in the gas density may 
cause an increase in the line ratio: If the gas in the dust lane is diffuse, that is, in the form of non-selfgravitating clouds, the lower critical density 
favors the emission of $^{\rm 12}$CO\,1$-$0 photons. As a result of the so-called radiative trapping the $^{\rm 12}$CO line is still bright at critical 
densities of $\sim$200-300~cm$^{\rm -3}$, where $^{\rm 13}$CO\,1$-$0 is faint \citep{mei04}. Previous studies of nearby galaxies have suggested that the 
diffuse gas in the dust lane is due to the funneling of gas along the same, thus gas infall is involved \citep[e.g.,][]{aalto00}. If the infalling gas is the 
only effect causing the increase of $\mathcal{R_{\rm 10}}$, the abundance ratio over the central region would need to be well mixed, that is, it should have 
values more corresponding to what is found for the inner Galactic disk \citep{mei04}. $\mathcal{R_{\rm 10}}$ around 30 at the center of NGC~1614 places the 
line ratios for the molecular gas there firmly above this \citep[see also Sect.\,\ref{subsec:12co-13co_line_ratio}; e.g.,][]{aalto95,mei04,aalto10}.\\
\indent
One possible candidate for a secondary line ratio enhancement mechanism could be a density wave in NGC~1614. The situation in 
\object{M~51} seems to be a good analogy for what we find in NGC~1614. In M~51, \citet{tos02} found high $^{\rm 12}$CO-to-$^{\rm 13}$CO\,1$-$0 ratios in 
the central and interarm regions. Velocity dispersion observations led them to suggest the presence of streaming motions. The authors thus proposed that 
streaming motions, caused by density wave activity, led to the accumulation of dense gas located in self-gravitating clouds, as traced by 
$^{\rm 13}$CO\,1$-$0, to be located downstream from the diffuse gas that is traced by $^{\rm 12}$CO\,1$-$0. This is exactly what we find in NGC~1614: 
the $\mathcal{R_{\rm 10}}$ is high in the dust lane where we suspect the gas to be diffuse. Downstream of this gas, traced by $^{\rm 12}$CO\,1$-$0, lies 
the peak of the $^{\rm 13}$CO\,1$-$0 distribution. This is also similar to what has been proposed for the Medusa merger \citep{aalto10}.

\subsubsection{Abundance effects} \label{subsubsec:abundance}

\begin{figure*}[h]
  \begin{minipage}[hbt]{0.4925\textwidth}
  \centering
    \includegraphics[width=0.75\textwidth]{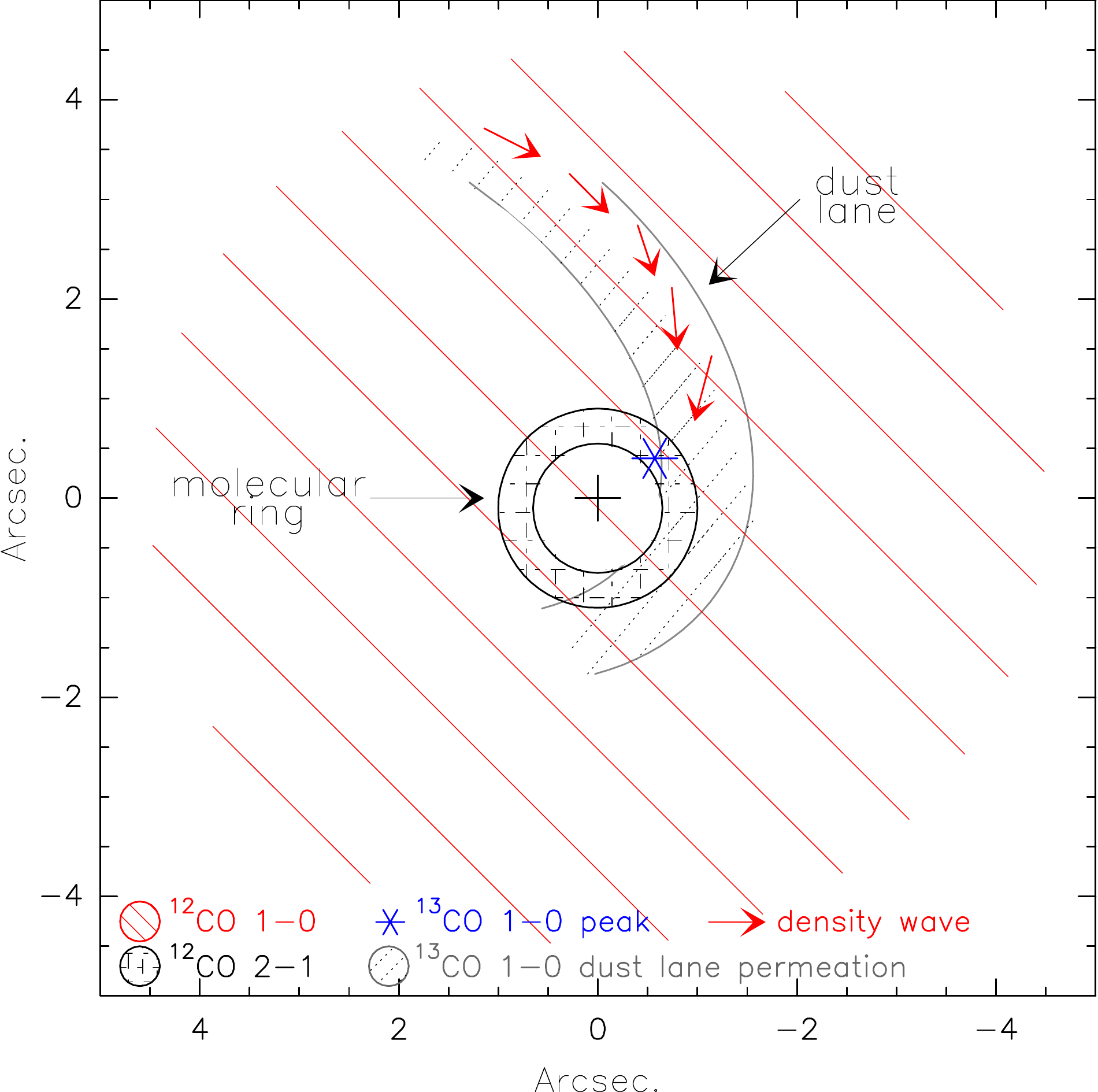}
  \end{minipage}
  \begin{minipage}[hbt]{0.4925\textwidth}
  \centering
    \includegraphics[width=0.75\textwidth]{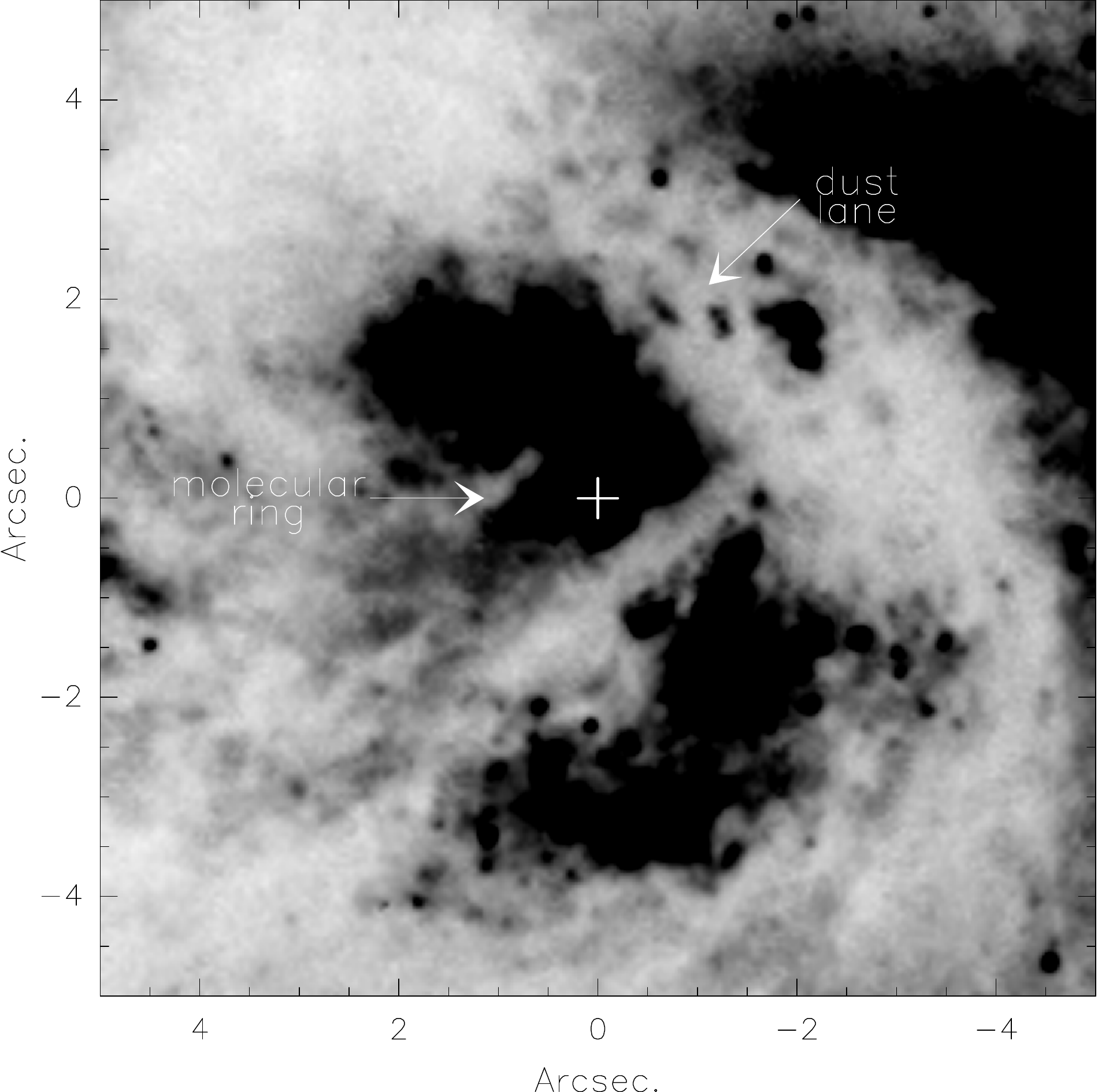}
  \end{minipage}
  \caption{\footnotesize Cartoon representation of the molecular gas structures in NGC~1614 (\textit{left}) and an \textit{HST} image of the same region 
   for comparison (\textit{right}). A high percentage of the $^{\rm 12}$CO\,1$-$0 emission (hatched red) is associated with the dust lane. The gas in the 
   dust lane is in the form of diffuse gas. The $^{\rm 13}$CO\,1$-$0 emitting gas is not fully distributed throughout the dust lane (hatched gray). Its 
   emission peak (blue star) is located in the ``umbilical cord'', the connection between the molecular gas in the circumnuclear ring, traced by 
   $^{\rm 12}$CO\,2$-$1 emission \citep[crosshatch pattern,][]{koenig13}, and the dust lane. The diffuse gas is being transported along the dust lane in 
   the form of unbound molecular clouds (GMCs). The interaction with density waves (red arrows) causes the gas in the clouds to become self-gravitating. In 
   the umbilical cord, crowding processes in the merger potential occur and the GMCs form GMAs through collisional coagulation. These larger associations 
   are then transported further onto the circumnuclear ring and star formation may be triggered.}
  \label{fig:big_picture}
\end{figure*}

\subsubsubsection{$^{\rm 12}$C/$^{\rm 13}$C} \label{subsubsubsec:abundance_12c_13c}

The $^{\rm 12}$CO-to-$^{\rm 13}$CO\,1$-$0 intensity line ratio map shows values of about 30 toward the nucleus of NGC~1614 (see 
Sects.\,\ref{subsec:12co-13co_line_ratio} and \ref{subsubsec:excitation}), which corresponds to line ratios found predominantly for warm, turbulent, 
high-pressure gas in the centers of luminous merging galaxies \citep[e.g.,][]{aalto95,gle01}. From the discussion of the excitation effects 
on the line ratios in Sect.\,\ref{subsubsec:excitation}, we conclude that diffuse gas in the dust lane under the influence of density 
wave activity can partially explain elevated values for $\mathcal{R_{\rm 10}}$. To increase the $^{\rm 12}$CO-to-$^{\rm 13}$CO\,1$-$0 to a level as we 
find in NGC~1614 an additional factor has to be taken into account to explain the enrichment of $^{\rm 13}$CO, which also causes the relative deficiency 
of C$^{\rm 18}$O. This is the effect of changing abundances.\\
\indent
\citet{cas92} suggested that $\mathcal{R}_{\rm 10}$ could change depending on the $^{\rm 12}$CO/$^{\rm 13}$CO abundance ratio, 
[$^{\rm 12}$CO]/[$^{\rm 13}$CO], when low-metallicity gas was transported from the outskirts of a merger to its center. The starburst would then contribute 
to enrich the infalling metal-poor gas in $^{\rm 13}$C \citep[e.g.,][]{rup08,hen10}. We assume the $^{\rm 12}$CO\,1$-$0 line transition to be optically 
thick ($\tau$\,=\,1), that the $^{\rm 12}$CO and $^{\rm 13}$CO emission trace the same gas and also that the excitation temperatures for the two are 
comparable\footnote{For more details see Sect.\,\ref{sec:appendix_2}}. This would mean that [$^{\rm 12}$CO]/[$^{\rm 13}$CO] would be roughly 90, which is a 
typical value for gas farther out in the Galactic disk \citep[e.g.,][]{hen85,wil94,wou96,hen14}, and thus points to the presence of infalling, chemically 
less processed gas.

\subsubsubsection{$^{\rm 16}$O/$^{\rm 18}$O} \label{subsubsubsec:abundance_16o_18o}

Using the C$^{\rm 18}$O\,1$-$0 integrated intensity upper limit in the central 4\arcsec\,$\times$\,4\arcsec, the 
$^{\rm 12}$C$^{\rm 16}$O-to-$^{\rm 12}$C$^{\rm 18}$O\,1$-$0 line ratio results in a lower limit of $\sim$325. The resulting abundance ratio is 
[$^{\rm 12}$CO]/[C$^{\rm 18}$O]\,$\sim$900. Values found for the Galactic center are $\sim$250, for the solar neighborhood typical values are 
about 500 \citep[e.g.][]{wil92}. The [$^{\rm 12}$CO]/[C$^{\rm 18}$O] ratio is significantly higher in NGC~1614 than in the LIRG \object{Zw~049.057} 
\citep{fal15}, but comparable to what has been found for another LIRG, \object{NGC~4418}, \citep{gon12}. It is also comparable to values found in the 
merger \object{Arp~299} (Falstad et al., in prep.). Interestingly, both Arp~299 and NGC~4418 have molecular gas inflows 
\citep[Falstad et al. in prep.,][]{cos13} -- Arp~299 also hosts an efficient starburst \citep[e.g.,][and references therein]{bon12} -- just like NGC~1614. 
Considering that this result for NGC~1614 is only a lower limit to [$^{\rm 12}$CO]/[C$^{\rm 18}$O], and also 
taking the high $^{\rm 13}$CO-to-C$^{\rm 18}$O\,1$-$0 integrated intensity line ratio in the central 4\arcsec\,$\times$\,4\arcsec into account 
($\sim$10.3), C$^{\rm 18}$O\,1$-$0 seems very deficient in comparison to $^{\rm 12}$CO\,1$-$0, but especially with respect to $^{\rm 13}$CO\,1$-$0. 
C$^{\rm 18}$O is thought to come from short-lived massive stars early during a starburst event \citep[e.g.,][]{pra96,mei04}, whereas 
$^{\rm 13}$CO\,1$-$0 is predicted to be produced later on in intermediate-mass stars \citep[e.g.,][and references therein]{mei04}. An explanation for 
this deficiency in the center of NGC~1614 could be that the mixing of the infalling, chemically unprocessed gas with the prevailing gas is not 
efficient. Although this could also be an effect of the size of the beam in the observations - if the unprocessed infalling gas is located inside the 
same beam as the prevailing nuclear gas a dilution of the separate signals could occur. This dilemma will be solved by higher resolution observations 
of the same set of molecular tracers.\\
\indent
Taking the findings in Sect.\,\ref{subsubsec:excitation} and this section into account, we propose that the increase in the 
$^{\rm 12}$CO-to-$^{\rm 13}$CO ratio in NGC~1614 is caused by the diffuse gas in the dust lane and density wave activity, in 
combination with enrichment of $^{\rm 13}$CO due to infalling metal-poor gas from farther out in the galactic disk. However, a definitive answer to 
whether a change in the temperature of the gas also has an influence can only be obtained together with sensitivity-matched observations of the 
$^{\rm 12}$CO and $^{\rm 13}$CO\,2$-$1 lines.

\subsection{NGC~1614 - the big picture?} \label{subsec:big_picture}


In this section, we now collocate our results together with what has been previously reported on what is going on in NGC~1614. A cartoon 
representation of the proposed scenario is depicted in Fig.\,\ref{fig:big_picture}.\\
\indent
In 2001, a starburst ring was discovered at the nucleus of NGC~1614 by \citeauthor{alo01}. Its presence was confirmed in a number of other tracers 
\citep[e.g.,][]{ols10,vai12}. \citet{alo01} suggested that the ring was formed as the result of an outward progressing starburst event that has 
already consumed most of the gas at its center. The ring also has a molecular component \citep{koenig13,sli14,xu15}. As a result, it was suggested that 
the ring is situated at the location of crowded orbits in the merger potential where it is replenished by gas coming in along the dust lane 
\citep{koenig13,sli14}. The $^{\rm 12}$CO\,1$-$0 \citep{ols10,sli14,gar15}, 2$-$1 \citep{koenig13} and $^{\rm 13}$CO\,1$-$0 line transitions trace the 
low-to-intermediate surface brightness gas that shows how the starburst ring is connected to the large-scale molecular gas reservoir: diffuse gas in 
the form of unbound giant molecular clouds (GMCs) could be funneled along the dust lane from the molecular gas reservoir at larger scales (seen in 
$^{\rm 12}$CO\,1$-$0) toward the nucleus. During this process, the gas may be hit by a density wave. The resulting shocks could cause the gas density to 
increase and the molecular clouds to become self-gravitating (as traced by $^{\rm 13}$CO\,1$-$0). In this scenario, the molecular clouds are trapped 
at the connection between the dust lane and the ring, ``the umbilical cord'' \citep[described by][]{koenig13}, inside the mergers potential through 
crowding processes, for instance. This could lead to collisional coagulation of the GMCs into larger-sized giant molecular associations (GMAs, traced by, 
e.g., $^{\rm 12}$CO\,2$-$1) that then possibly migrate onto the circumnuclear ring. In this way, the nuclear gas reservoir can be constantly replenished. 
The cloud-cloud interactions could also trigger the onset of star formation in the ring \citep{sco86,tan00}: the high-density gas, such as HCN and 
HCO$^{\rm +}$ \citep[e.g.,][]{ima13,sli14,xu15}, is exclusively associated with the star formation itself.\\
\indent
Whether the ring is indeed caused by a wildfire expanding into the surrounding molecular medium or if is solely formed at the location of the resonance in 
the merger potential is still under discussion. Solving this argument will require further studies: Resolution- and sensitivity-matched $^{\rm 12}$CO and 
$^{\rm 13}$CO\,2$-$1 observations will greatly improve our ability to determine whether the presence of non-selfgravitating gas in the dust lane or a 
change in the temperature of the gas is the cause of the extreme line ratios in NGC~1614. Higher resolution $^{\rm 12}$CO\,1$-$0 observations at high 
sensitivity are needed to search for indicators of possible streaming motions between the large-scale and small-scale molecular gas reservoirs to verify 
the funneling of gas along the dust lanes toward the circumnuclear ring. They would also allow us to study the possible connection between the CO\,1$-$0 
and the CO\,2$-$1 gas reservoirs in more detail, thus allowing to collect evidence in favor or against the competing scenarios proposed for the formation 
of the circumnuclear ring.

\subsubsection{Implications for high-redshift studies of mergers} \label{subsubsec:high-z}

We would also like to stress that our results emphasize the need to be cautious which tracers of the molecular gas to best use to determine star 
formation capabilities, also for studies at high redshifts. We have especially shown in this paper that the $^{\rm 12}$CO\,1$-$0 emission is not 
necessarily a tracer of the overall content of the molecular gas that is capable to partake in star formation because a high percentage of it might 
be in the form of diffuse molecular gas. Additional observations at higher-$J$ CO transitions are necessary to conclude on this. Furthermore, 
C$^{\rm 18}$O should be used with caution to determine the age of the starburst activity in galaxies with proven gas infall. The mixing of the 
infalling gas with the prevailing gaseous materials might falsify the age determination to a large degree.


\section{Summary} \label{sec:summary}

In summary, with the ALMA observations presented here, we showed that the nearby starburst galaxy NGC~1614 harbors large reservoirs of molecular gas 
traced by $^{\rm 12}$CO and $^{\rm 13}$CO\,1$-$0. The $^{\rm 12}$CO emission is widely distributed throughout the galaxy and has a strong connection 
to the dust lanes, whereas the $^{\rm 13}$CO emission is much more compact and seems to avoid them. This possibly indicates non-selfgravitating, 
diffuse gas in the dust lane that originates from farther out in the galaxy disk. The $^{\rm 13}$CO\,1$-$0 emission distribution is most likely a result 
of the effect of the progression of density waves on the galaxies molecular medium. In addition to $^{\rm 12}$CO and $^{\rm 13}$CO\,1$-$0, 
other molecular gas tracers such as CN\,1$-$0 and CS\,2$-$1 are present in NGC~1614. For C$^{\rm 18}$O only an upper limit was found. 


\begin{acknowledgements}
      This paper makes use of the following ALMA data: ADS/JAO.ALMA\#2013.1.00991.S. ALMA is a partnership of ESO 
      (representing its member states), NSF (USA) and NINS (Japan), together with NRC (Canada), NSC and ASIAA (Taiwan), and KASI (Republic of 
      Korea), in cooperation with the Republic of Chile. The Joint ALMA Observatory is operated by ESO, AUI/NRAO and NAOJ. This research has 
      made use of the NASA/IPAC Extragalactic Database (NED) which is operated by the Jet Propulsion Laboratory, California Institute of 
      Technology, under contract with the National Aeronautics and Space Administration.
\end{acknowledgements}


\bibliographystyle{aa}
\bibliography{ngc1614_alma}

\begin{appendix}

\section{Comparison overlays of $^{\rm 12}$CO\,1$-$0, 2$-$1 and $^{\rm 13}$CO\,1$-$0 emission distributions} \label{sec:appendix_1}

\begin{figure}[h]
  \centering
    \includegraphics[width=0.4\textwidth]{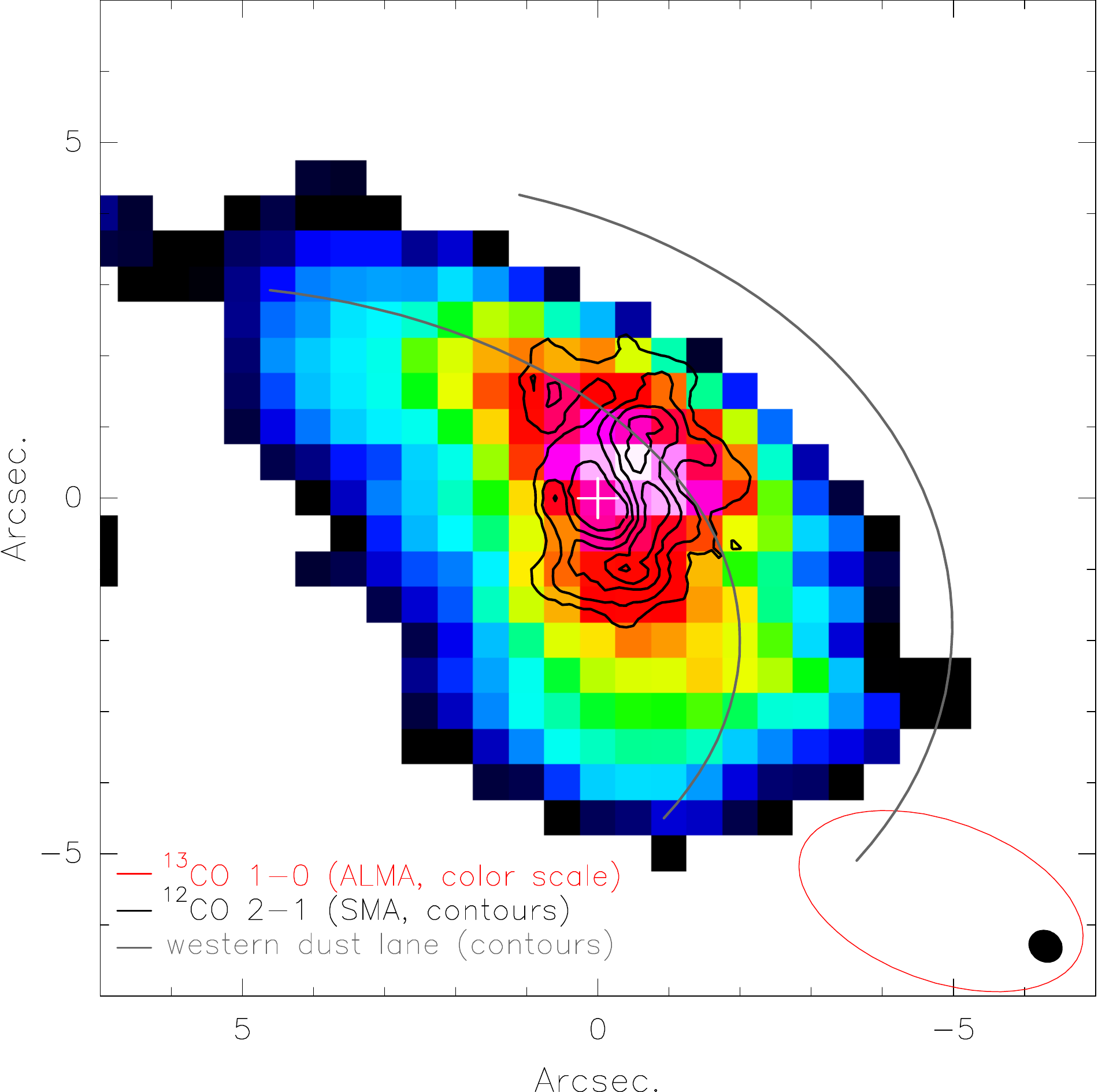}
  \caption{\footnotesize Comparison of high-resolution $^{\rm 12}$CO\,2$-$1 observations with the SMA \citep[black contours,][]{koenig13} to the ALMA 
   $^{\rm 13}$CO\,1$-$0 emission presented in this work (color scale). Note how the $^{\rm 13}$CO\,1$-$0 emission peaks exactly at the location of the 
   connection between the molecular ring and the dust lane \citep[grey contours, see also Figs. 1a \& 3 in][]{koenig13}. The $^{\rm 12}$CO\,2$-$1 
   contours start at 4$\sigma$ and are spaced in steps of 4$\sigma$ (1$\sigma$ = 6.16~mJy\,beam$^{\rm -1}$). North is up, east to the left, and the beams 
   are shown in the lower right corner of the image. The cross marks the position of the phase center of the observations.}
  \label{fig:overlay_12co2-1_on_13co1-0}
\end{figure}

\begin{figure}[h]
  \centering
    \includegraphics[width=0.4\textwidth]{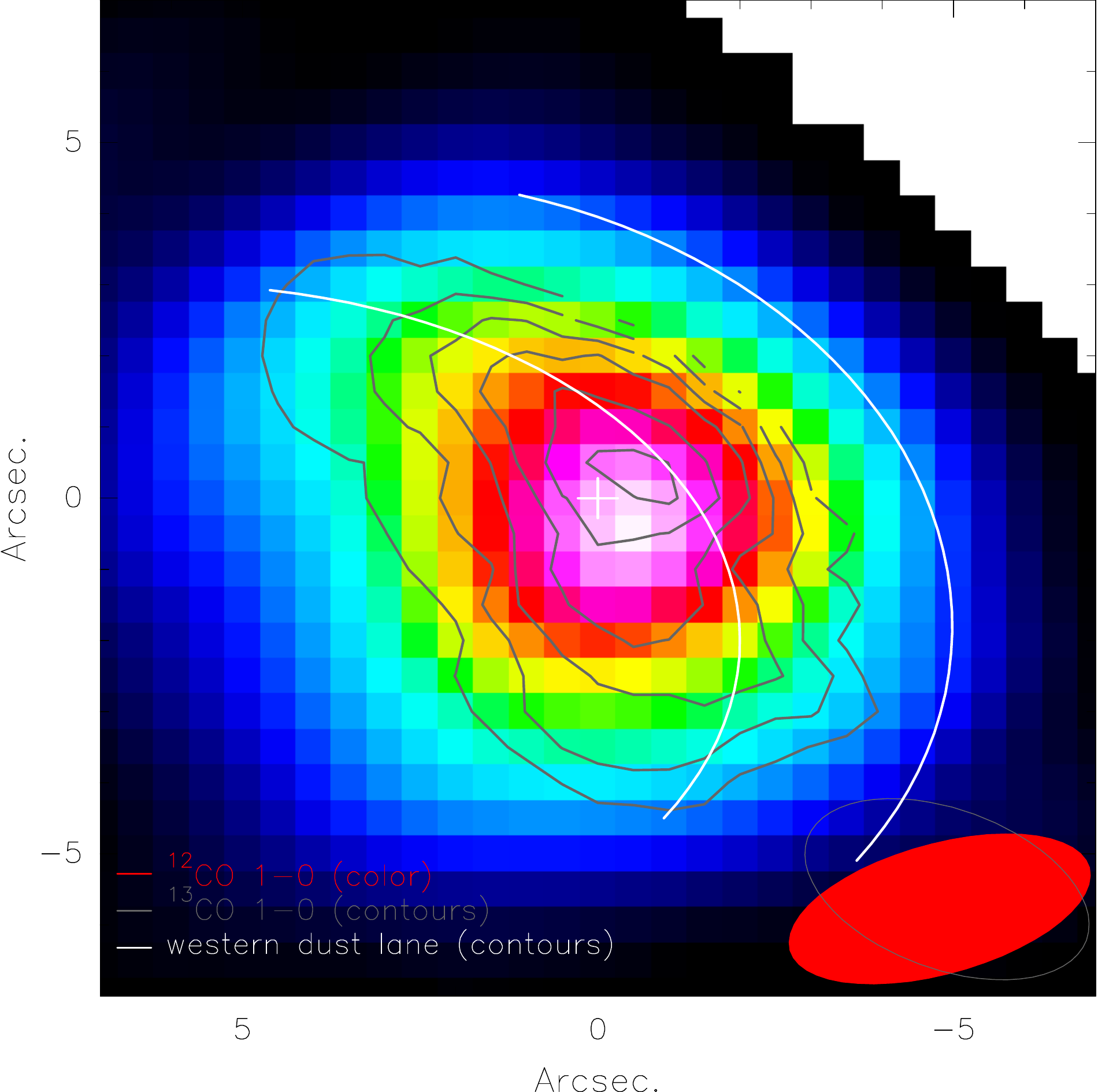}
  \caption{\footnotesize Overlay of the $^{\rm 13}$CO\,1$-$0 emission observed with ALMA (gray contours) on top of the $^{\rm 12}$CO\,1$-$0 emission 
   observed with ALMA (color scale). Note how the bulk of the $^{\rm 13}$CO\,1$-$0 emission is located outside the dust lane to the west of the nucleus 
   (white contours). Moreover, the $^{\rm 13}$CO\,1$-$0 located in the dust lane does not cover its full extent. The $^{\rm 13}$CO\,1$-$0 contours 
   are the same as in Fig.\,\ref{fig:overlay_13co_false_color}b. North is up, east to the left, and the beams are shown in the lower right corner of the 
   image. The cross marks the position of the phase center of the observations.}
  \label{fig:overlay_13co1-0_on_12co1-0}
\end{figure}

Figures \ref{fig:overlay_12co2-1_on_13co1-0} and \ref{fig:overlay_13co1-0_on_12co1-0} show overlays of $^{\rm 12}$CO\,1$-$0, 2$-$1, and 
$^{\rm 13}$CO\,1$-$0 that might be useful for a closer look at the detailed structure of NGC~1614. The observations used here are presented in this 
work or have been previously published.

\section{Conversion of line ratios into abundance ratios} \label{sec:appendix_2}

To study the effects of changes in abundance ratios vs. the effects of different excitation mechanisms, we used the $^{\rm 12}$CO-to-$^{\rm 13}$CO\,1$-$0 
ratio distribution in NGC~1614 to derive the abundance ratio ([$^{\rm 12}$CO]/[$^{\rm 13}$CO]). While $\tau$($^{\rm 12}$CO\,1$-$0) in galaxy disks 
(i.e., an ensemble of GMCs) may well exceed unity, there is mounting evidence that the emitting CO surfaces in galaxy centers have moderate optical depths 
\citep[e.g.,][]{aalto95,dow98,isr09}, this is also true for the Galactic center \citep[e.g.,][]{pol88,dah98}. Since a $\tau$\,=\,1 surface would have the 
optical combination of filling factor and brightness temperature it is a reasonable assumption that much of the CO emission would be emerging from these 
surfaces. Thus, we use as a first-order approximation of the optical depth of $\tau$\,=\,1 in our data as a starting point to study the physical gas 
properties and mechanisms with a simple model. This is justified for the molecular gas in NGC~1614 based on several pieces of evidence: 1) In the dust 
lane the gas density is low: Studies associate high-density gas tracers exclusively with the circumnuclear ring \citep[e.g.,][]{ima13,sli14,xu15}; thus, 
the brightness temperature measured in the dust lane suggests that the gas has a $\tau$ of $\geq$1. If the emission is optically thin in a cool, low-density 
environment, its brightness temperature is $<$1~K at this resolution. 2) In the dense and warmer gas in the center \citep[e.g.,][]{ima13,sli14,xu15}, even 
optically thin $^{\rm 12}$CO\,1$-$0 may have T$_{\rm B}$\,$<$1~K. However, the $\tau$\,$\geq$1 surfaces would contribute more to the total intensity and 
$^{\rm 12}$CO is bathed in an intense field in the starburst region and cannot self-shield until $\tau$\,=\,1 \citep[e.g.,][]{vanD88,vis09}. The high 
$^{\rm 13}$CO-to-C$^{\rm 18}$O ratio supports this. In NGC~1614, the gas with the highest $\mathcal{R_{\rm 10}}$ ratios is located in the dust 
lane, away from the bulk of the ongoing star formation, and thus away from the most intense radiation. Therefore, the intense radiation field from the 
star formation in the ring and/or self-shielding of $^{\rm 12}$CO are unlikely to cause the high $\mathcal{R_{\rm 10}}$ (and the high isotopic ratio) in 
the western dust lane.\\
\indent
More higher resolution studies of the CO\,1$-$0 isotopologs are necessary to set a tighter constraint on the optical depth structure of the molecular gas 
in NGC~1614.

\end{appendix}

\end{document}